\documentclass[fleqn,usenatbib]{mnras}

\usepackage{newtxtext,newtxmath}

\usepackage[T1]{fontenc}
\usepackage{ae,aecompl}

\usepackage{graphicx}	
\usepackage{amsmath}	

\title[Planetary Embryo Collisions]{Planetary Embryo Collisions and the Wiggly Nature of Extreme Debris Disks}

\author[L. Watt et al.]{Lewis Watt$^{1}$\thanks{E-mail: lewis.watt@bristol.ac.uk},
Zoe Leinhardt$^{1}$,
Kate Su$^{2}$
\\
$^{1}$School of Physics, University of Bristol, H.H. Wills Physics Laboratory, Tyndall Avenue, Bristol BS8 1TL, UK\\
$^{2}$Steward Observatory, University of Arizona, 933 North Cherry Avenue, Tucson, AZ 85721, USA
}

\date{Accepted XXX. Received YYY; in original form ZZZ}

\pubyear{2020}

\begin{document}
\label{firstpage}
\pagerange{\pageref{firstpage}--\pageref{lastpage}}
\maketitle

\begin{abstract}
In this paper, we present results from a multi-stage numerical campaign to begin to explain and determine why extreme debris disk detections are rare, what types of impacts will result in extreme debris disks and what we can learn about the parameters of the collision from the extreme debris disks. We begin by simulating many giant impacts using a smoothed particle hydrodynamical code with tabulated equations of state and track the escaping vapour from the collision. Using an $N$-body code, we simulate the spatial evolution of the vapour generated dust post-impact. 

We show that impacts release vapour anisotropically not isotropically as has been assumed previously and that the distribution of the resulting generated dust is dependent on the mass ratio and impact angle of the collision. In addition, we show that the anisotropic distribution of post-collision dust can cause the formation or lack of formation of the short-term variation in flux depending on the orientation of the collision with respect to the orbit around the central star. Finally, our results suggest that there is a narrow region of semi-major axis where a vapour generated disk would be observable for any significant amount of time implying that giant impacts where most of the escaping mass is in vapour would not be observed often but this does not mean that the collisions are not occurring.
\end{abstract}

\begin{keywords}
circumstellar matter -- planets and satellites: formation -- method: numerical
\end{keywords}



\section{Introduction}
\label{section:intro}
Energetic impacts between planetary embryos are common in the end stages of terrestrial planet formation \citep{Agnor_1999_Large_impacts,O'Brien_2006_Terrestrial_planet_formation,Morishima_2010_planetesimals_to_terrestrial_planets,Elser_2011_How_common_Earth_Moon_systems}. After the dissipation of the gas in a protoplanetary disk, the terrestrial planet region is populated by planetesimals and planetary embryos. Without the eccentricity damping from the protoplantary gas, gravitational interactions between planetary embryos can lead to frequent orbit crossing. The added dynamical excitation causes the terrestrial region to become a hotbed of collisions in which large scale, giant impacts can occur \citep{Chambers1998-Gasremoved-cols}. It is clear that giant impacts played an important role in the formation of the terrestrial planets in our own Solar System, for example, the Earth-Moon system is almost certainly the result of a giant impact \citep{Hartmann_&_Davis_1975_Lunar_origin,Canup2004-LunarFormation,Cuk_&_Stewart_2012_Moon_fast_spinning_Earth,Canup_2012_Forming_Moon_Earth_like_comp} and the high core to mantle ratio of Mercury also suggests a large impact event \citep{Benz-2007-origin-of-mercury}. Giant impacts might also contribute to the formation of some Super-Earths as suggested by \citet{Bonomo2019_Kepler107impact}. 
While there is evidence of large impacts having played a role in the formation of the solar system, the lack of direct observations of planet formation in exoplanetary systems makes it difficult to compare theory and numerical simulations to reality.  

Traditional debris disks are typically exo-Kuiper belts inhabiting regions relatively far from the host star. The long orbital timescale results in slow collisional evolution that can last for  hundreds of millions of years. Irradiation of the fine, micron-sized dust within the debris disk produces excess infrared emission above what is expected from the central star alone. The dust is replenished through a steady-state collisional cascade, in which a background, unseen, population of boulders/planetesimals is ground down \citep{Wyatt-2008-evolution-of-debris-disks}. Dust is lost over time due to processes such as Poynting-Robertson drag or the ejection of particles smaller than the blow-out size through radiation pressure.  The lifetime of the disk is, therefore, set by the collisional activity between the planetesimals, which is dependent on the radial location, the total mass of colliding bodies and the size of the largest objects within the disk.

Because debris disks traditionally are associated with collisional erosion of planetsimals (a relatively slow process) they are generally treated as a steady state. Other dynamical activities such as extra perturbations due to newly formed planets or on-going planet accretion through collisions can also potentially produce additional source material for a dusty observable debris disk \citep{Dobinson_2016}.  
Particularly, most giant impacts are not perfect merging events even if they result in substantial growth of the target. Often collisions produce a range of escaping material from a gravity dominated boulder population (1-100 km) \citep{Leinhardt-2012-collisions-between-gravity-dom-bodies} to small dust grains formed from vapour condensate \citep{Johnson2012-vaporplumes}. A Moon-forming giant impact at 1 au would be observable at 25 $\mu$m for about $\sim$24 Myr assuming a standard collisional cascade in a boulder population with a maximum radius of 500 km \citep{Jackson2012_Moon}. 
Furthermore, the predicted observing signature of a giant impact would also depend on the detailed treatment of collisions within the fragments, but generally expect to last for over 1000 orbital periods \citep{Jackson-2014-planetary-collisions-at-large-au}. However, recent observations of extreme debris disks, a subclass of debris disks that are relatively young, bright and warm \citep{rhee_2008,Melis_2010,Meng2012-variability}, have found that some show significant variability and flux decline on a much shorter timescale \citep{Melis_2012,Meng2014-ID8science}, indicating that a standard steady-state collisional cascade might not be the correct explanation.

\subsection{Previous Work on the ID8 and P1121 Extreme Debris Systems}

Extreme debris disks, such as ID8 \citep{Meng2012-variability,Meng2014-ID8science,Su-2019-extreme-disk-variability} and P1121 \citep{Meng2015-pc,Su-2019-extreme-disk-variability}, typically exist around stars with ages between 10 and 200 Myr, and have a fractional luminosity, $f=L_d/L_* > 10^{-2}$, which is much brighter than traditional debris disks, $f<10^{-4}$, hence the term "extreme". They are populated by warm dust ($\geq 150$ K), hotter than a traditional debris disk ($< 150$ K), suggesting the presence of dust close to the star within the terrestrial zone. Extreme debris disks, therefore, exist at a time when the collisional activity in the terrestrial zone is at its greatest and the large luminosities indicate a significant influx of small dust particles. A possible source for a sudden abundance of small dust is from condensates of vapourised silicates ejected from a recent giant impact. 

ID8 is one of the most well studied extreme debris disks with its well documented disk light curves showing consistent but unpredictable variability. ID8 is a solar-like star (G6V and solar metallicity) located in the 35 Myr old open cluster NGC 2547. Its mid-IR spectrum taken in 2007 showed strong silicate features between $8\ \rm{\mu m}$ and $30\ \rm{\mu m}$, suggesting very fine dust surrounding the star. The disk has a fractional luminosity of $f=3.2\times 10^{-2}$ and a temperature of $\sim 750 \rm{K}$, putting the dust disk within the terrestrial planet formation zone which could extend out to 2-3 au. At the beginning of 2013, observations of ID8 using Spitzer IRAC camera at $3.6\rm{\mu m}$ and $4.5\rm{\mu m}$ showed a large increase in the flux compared to the early flux in 2012, indicating a sudden increase of dust. The leading hypothesis given by  \citet{Meng2014-ID8science} was a giant impact between two embryos during the unobserved period between 2012 and 2013. During the subsequent evolution in 2013, the overall flux rapidly decreased over a timescale of $\sim 370$ days. The observed depletion timescale is at least two orders of magnitude smaller than that calculated using a typical collisional cascade \citep{Meng2012-variability}. 

Intriguingly, ID8 also displayed another increase in excess infrared flux from 2014 to 2015, with a sudden drop between 2015 and 2016 \citep{Su-2019-extreme-disk-variability}.
These rapid increases and decreases in flux can not be explained by the slow grinding of kilometer scale planetesimals -- the timescales of the observed variability is too short. 
However, it is possible to produce the observed flux decline with the collisional evolution of small grains from condensation of vapour produced by energetic giant impacts between planetary embryos. Vapour produced from impacts will condense into grains with a narrow size distribution, peaked around a characteristic size, which vary from $\rm{ 10\ \mu m}$ to mm/cm depending on the collision parameters \citep{Johnson2012-vaporplumes}. A debris disk fed by underlying grains of 1 mm maximum size will have a significantly shorter collisional cascade lifetime than a traditional debris disk supplied by a kilometer scale boulder population. Assuming the vapour condenses into a characteristic small size it is possible to reproduce the 2013 flux decline observed in ID8 with a collisional cascade \citep{Meng2014-ID8science}.

ID8, while varying significantly in IR flux on an annual timescale, also shows a complex short-term flux modulation in the 2013 and 2014 data segments \citep[Fig. 2 and 3]{Su-2019-extreme-disk-variability}. One explanation for the short-term variability was proposed by \citet{Jackson-2014-planetary-collisions-at-large-au}. This work showed that dust ejected isotropically from the collision point will pass through two common locations along the orbit, namely, the collision point and anti-collision line. At these specific locations there will be an observable reduction in the flux as the surface area of the dust is decreased because all of the dust grains pass through both locations \citep{Jackson-2014-planetary-collisions-at-large-au}. If the viewing geometry of the system is close to edge on, as the dust passes through one of the disk ansae the column density of the dust will increase, leading to a larger optical thickness that also decreases the flux. The period of the oscillatory behaviour will depend on the inclination of the system and the impact location. Therefore, two identical collisions at different locations in a system will produce different short-term ocillatory periods in flux. This is seen in ID8: the 2013 data suggest an orbital period of 108 days for the dust clump while the 2014 data indicate an orbital period of 41.6 days, which translate to semi-major axis of the dust clump/collision to be at 0.43 au and 0.24 au, respectively.

P1121 is another system similar to ID8. Located in the 80 Myr old open cluster M47, with spectral type F9V, it has been estimated to have a debris disk with a fractional luminoisity of $\sim 2\times 10^{-2}$ \citep{Meng2015-pc}. {\it Spitzer} 3.6/4.5 $\mu$m data since 2012 suggest a gradual flux decline with a timescale of 310$\pm$60 days, reaching to a background level since 2015 \citep{Su-2019-extreme-disk-variability}. On top of the flux decline, a weak modulation with an apparent periodicity of 18 days is also seen.  The true orbital period of the clump would either be 36 or 72 days if the modulation is caused by the optically thick clump. 

The detection rate of observing extreme debris disks does not match the rate that is required to explain the frequency of terrestrial planets \citep{Meng2015-pc}, with hot dust only being found around $1\%$ of young Sun-like, FGK, stars \citep{Kennedy_Wyatt_2013_bright_end_of_exozodi_luminosity}.  This points towards one of two possibilities: 1) that large embryo-embryo collisions are not as prominent in the planet formation process as models predict or 2) the post-collision escaping material does not survive and is thus not observable for as long as predicted. 

\subsection{Aims}

With nearly all current models of terrestrial planet formation involving giant collisions between planetary embryos, the low incident of extreme debris disks is a large problem. Previous work has been able to explain some of the general flux behaviour  \citep{Jackson2012_Moon,Jackson-2014-planetary-collisions-at-large-au,Kral_2015_dust_evolution} and the oscillatory behaviour shown in some disks \citep{Su-2019-extreme-disk-variability} by assuming that the escaping material after a giant collision is isotropically distributed. However, the timescale that a clump shears out due to its Keplerian motion is dependent on the initial velocity distribution of escaping debris, meaning an isotropic distribution may not be realistic. The asymmetry in a given collision can have a profound effect on the resulting debris disk. In this paper we have conducted a wide range of giant impacts to better understand how the distribution of escaping vapour will differ between various impact parameters with the goal of understanding the effects of said impacts on detection rate of extreme debris disks. We also look at the initial evolution of the escaping material for the whole suite of giant impact simulations to find what causes the very short term oscillatory behaviour in some extreme debris disks.
Our methods are outlined in section 2, which explains each step of our numerical investigation from the simulations of giant impacts using smoothed particle hydrocode (SPH), to the evolving of the dust produced using a basic $N$-body integrator, and finally to the radiative transfer calculations of the debris emission by RADMC3D. In section 3, we discuss the observability of extreme debris disks and how it will vary with differing parameters. Section 4 outlines the limitations of our study. We conclude our results in section 5.

\section{Methods}
\label{section:methods}
Our numerical method is divided into three steps: 1) direct SPH simulations of giant impacts; 2) $N$-body simulations of escaping post-collision vapour; and 3) radiative transfer and production of a synthetic observation.

\subsection{Step 1: Giant Impacts (SPH)}
We model planetary embryo-embryo collisions using a modified version of the SPH code Gadget 2 \citep{Springel-2005-Gadget-2,Marcus2009-ANEOS}. For source code see \citealt{Cuk_&_Stewart_2012_Moon_fast_spinning_Earth}. This modified version uses tabulated equations of state (EOS) to determine the thermodynamic properties of materials. The planetary embryos are a composite of iron (core) and forsterite (mantle) with a mass ratio of 3:7 and given initial temperature profiles from  \citet{Valencia-2005-internal-structure}. Both core and mantle use the ANEOS equations of state \citep{Melosh2007-ANEOS,Marcus2009-ANEOS} and are initialised and equilibrated similar to previous work \citep{Carter2018,Denman2020}.

A summary of all collision simulations presented in this work is shown in Tables \ref{tab:collisions_less_1M_e} and \ref{tab:collisions_geq_1M_e} in the Appendix. In these simulations impact speeds are varied from 6 $\rm{km\ s^{-1}}$ (just below the velocity needed to vaporise forsterite, \citealt{Davies_2020_collision_vapour_speed}) to few $v_{esc}$ ($\sim 20 \rm{km\ s^{-1}}$), where 
\begin{equation}
v_{\rm{esc}} = \sqrt{\frac{2G\,(M_{\rm{targ}}+M_{\rm{proj}})}{R_{\rm{targ}}+R_{\rm{proj}}}},
\end{equation} 
is the mutual escape speed and $M_{\rm{targ}}$, $M_{\rm{proj}}$, $R_{\rm{targ}}$, and $R_{\rm{proj}}$ are the masses and radii of the target and projectile respectively. Note, the maximum speeds of simulations at higher impact angle were extended to $\sim 5 v_{esc}$. The range of impact speeds were chosen to be fast enough to produce some vapour but slow enough to result in either embryo growth or a hit-and-run event as informed by results from numerical simulations of planet formation  \citep[for example,][]{Carter_2019_collision_speeds}. Three different impact parameters of $b=0,\ 0.4,\ \rm{and}\ 0.8$ (where $b = \sin \theta$ and $\theta$ is the impact angle) ranging from head-on to grazing were used to test how the initial impact parameter influences the observability of the escaping material. Mass ratio ($\mu = \rm{M_{proj}}/M_{targ}$) is varied from 0.09 to 1 to understand if differing masses between the target and the projectile will have a large effect on the distribution of escaping material. The particle resolution used for the planetary embryos was between $2\times 10^4$ and $2\times 10^5$. This resolution allows for a large range of simulations to be run in reasonable time and is sufficient for determining the mass and distribution of the escaping vapour.

\begin{figure*}
    \centering
    \includegraphics[width=\linewidth]{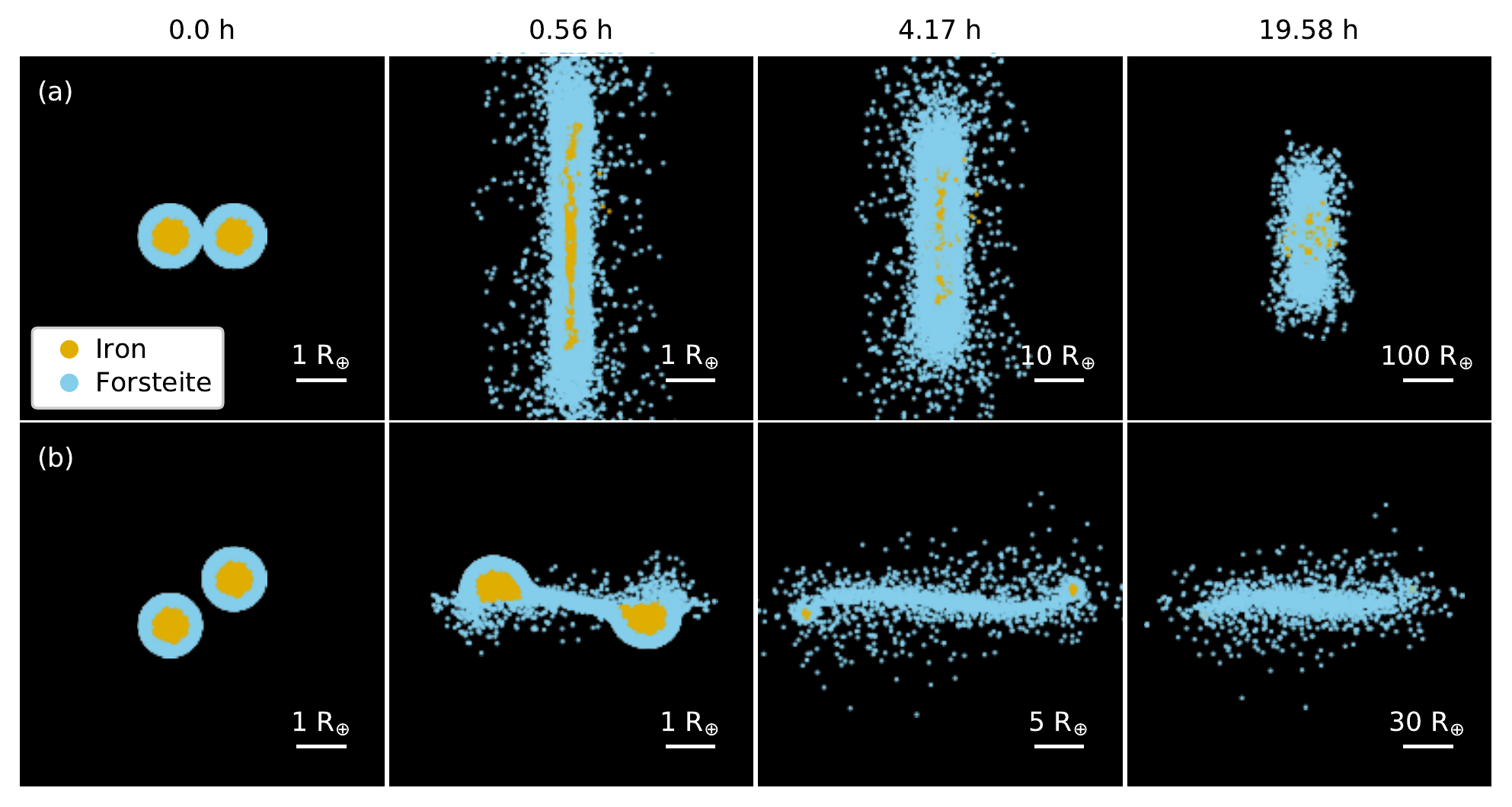}
    \caption{A hemispherical view of a giant impact between two planetary embryos of $0.1\rm{M_{\oplus}}$ at $10\rm{km\ s^{-1}}$ over 4 snapshots ranging in time from the initial collision to 19.58 hours post collision. A head-on impact (sim. 8, Table \ref{tab:collisions_less_1M_e}) is shown in (a) and a hit-and-run grazing collision (sim. 88, Table \ref{tab:collisions_geq_1M_e}) is shown in (b). Material type is indicated by colour: forsterite mantle in blue and iron core in yellow. Images scale with distribution size with scale bar shown in bottom right hand corner of each subfigure.}
    \label{fig:embryocol}
\end{figure*}

Even the relatively narrow range of collision parameters chosen for this work produce a large variety of collision outcomes. Figure \ref{fig:embryocol} shows four snapshots of two isolated giant impacts at $10\ \rm{km\ s^{-1}}$ (1.99 $v_{\rm{esc}}$) 
between equal mass 0.1 M$_\oplus$ embryos at two different impact parameters: (a) head-on ($b = 0$, sim 8 in Table \ref{tab:collisions_less_1M_e}) and (b) grazing collision ($b = 0.8$, sim 88 in Table \ref{tab:collisions_less_1M_e}). Both collisions show non-isotropic distribution of material after the impact and continues past $\sim$19 hours of simulation time. In simulation 8 one significant largest remnant is produced with a mass of  $1.45\times 10^{-1}\rm{M_{\oplus}}$ and an escaping mass of $4.54\times 10^{-2}\rm{M_{\oplus}}$. About 13.8\% of the escaping material is in the form of vapour. The grazing collision shows an example of a hit-and-run collision with two largest remnants of equal size ($M_{\rm{lr}} = 9.56\times 10^{-2}\rm{M_{\oplus}}$, roughly the initial mass of the targets). The amount of escaping mass is much less than the head on collision  ($M_{\rm{unb}} = 7.18\times 10^{-3}\rm{M_{\oplus}}$) and only 5.28\% of the escaping material is vapour. We would expect these two collisions to look very different observationally. Just how different is something we investigate in the rest of this paper. The simulation outcomes are determined by methods laid out in the next subsection.

\subsection{Step 2: Post-collision Dust Evolution ($N$-body)}
\label{section:post-collision_dust_evolution}

To determine the observability of the vapour generated dust we need to constrain the dynamical behaviour and lifetime of the observable dust post-collision, which are governed by two parameters: the mass of the largest remnant(s) and the spatial and velocity distributions of the escaping material. In order to do this we developed a $N$-body code using a leapfrog integrator with adaptive timestepping. The collision outcomes from our isolated SPH simulations become input for our $N$-body code. 

It is possible that some extreme debris disks actually show flux contributions from both the quickly generated and short lived vapour condensed dust and a secondary traditional debris disk generated by the boulder-sized giant impact remnants. In this paper we will only focus on the early vapour generated dust disk. In future work we will include the background boulder population and investigate the possible interaction between the two dust populations.

\subsubsection{Determining the Largest Remnant}

The first step in determing how much vapour escapes from a given giant impact is to determine the mass of the largest post-collision remnant. In this work an iterative process is used starting with the last snapshot (between 19 and 28 hours post collision). 

Each particle in the simulation is labelled as part of a resolved remnant, unresolved remnant, or identify it as unbound. The process begins by finding a seed particle. The seed particle is defined as the particle with the minimum potential energy when compared to all other particles,
\begin{equation}
    E_j = -\sum\limits_{i}^{N_{\rm{unb}}} \frac{Gm_im_j}{\lvert \mathbf{r_j} - \mathbf{r_i}\rvert},
\end{equation}
where $E_j$ is the potential energy of particle $j$, $m_i$ is the mass of particle $i$, $r_j - r_i$ is the separation between particle $j$ and $i$, and $N_{\rm{unb}}$ is the list of particles not including particle $j$, the seed particle is the particle that has the lowest $E_j$.

Particles are added to the seed to build up the remnant if the sum of their kinetic energy and potential energy is less than zero in the centre of mass frame of the remnant,
\begin{equation}
    E_k + E_p < 0,
\end{equation}
where the kinetic energy is
\begin{equation}
E_k = \frac{1}{2}\, m_iv^{\prime 2}, 
\end{equation}
$v^{\prime 2}= \lvert \mathbf{v_i} - \mathbf{v_b}\rvert^2$ and $\mathbf{v_b}$ is the velocity of the centre of mass of the remnant. The potential  energy is
\begin{equation}
E_p = -\frac{GM_{b} m_i}{r^{\prime}},
\end{equation}
where $M_b$ is the mass of the remnant and
$r^{\prime}=\lvert \mathbf{r_i} - \mathbf{r_b}\rvert =\sqrt{(\mathbf{r_i}-\mathbf{r_b})\cdot(\mathbf{r_i}-\mathbf{r_b})}$.

This iterative process will continue over all particles not bound to any remnant until the number of particles in the remnant has stabilised.
This process is then repeated until all remnants are found. Any particle not added to a remnant is labelled unbound. Remnants found with less than 500 particles are considered unresolved clumps.

\subsubsection{Finding the Vapour Fraction}

ID8 and P1121 show interesting behaviour, one of which is the decay of their infrared flux on the timescale of about a year (multiple instances demonstrated in the case of ID8). This decay cannot be attributed to the blow-out of small grains (too quick), or the loss of mass through a typical steady-state cascade of km-sized planetesimals (too slow). Currently the most likely hypothesis is that the excess flux is caused by vapour condensate which is created by a giant impact between planetary embryos. The vapour quickly condenses into grain sizes that can range from microns to mm/cm \citep{Meng2014-ID8science,Meng2015-pc,Su-2019-extreme-disk-variability}. The small characteristic size allows for a short collisional evolution. Therefore, for us to be able to simulate the early evolution of extreme debris disks, we need to know how much vapour is created in each impact and how it is distributed. 

In order to determine the amount of vapour the ANEOS equations of state are used to define a temperature-entropy vapour dome for each material. With the vapour dome and the lever rule, we can determine the mix of melt-to-vapour of each particle that sits inside the vapour dome. Particles are shocked onto the Hugoniot curve and then isentropically cool down to the triple point temperature, at which point both melt and vapour material will solidify forming spherules, droplets or dust \citep{Davies_2020_vapour}
. We use the lever rule at the triple point temperature to determine the vapour fraction of each particle, using $1890^{\rm{o}}$C as the triple point temperature for forsterite \citep{Nagahara1994-forsteritetp} and $2970^{\rm{o}}$C for iron \citep{Liu1975-TPIron}. After obtaining the mix of melt and vapour for each particle, we calculated the total escaping vapour mass by the summation of all the particle masses multiplied by the fraction of vapour mass they have. Our study ignores the melt mass as we assume this will play no part in the instantaneous visibility of a varying extreme debris disk, as melt forms droplets with a characteristic size a few orders of magnitude larger than vapour material \citep{Johnson2014-meltdroplets}, or larger objects like boulders.

\subsubsection{Dynamical Evolution}

After identifying the largest remnants and the vapour fraction mass we hand off the data to a $N$-body code. The system now needs to be evolved in time. 
In the $N$-body simulations, the star (1 M$_\odot$) is at the origin. The centre of the dust distribution and the largest post-collision remnants are placed at the collision point, where the centre of the distribution is defined as the centre of mass of the largest remnant(s) of the isolated SPH simulation. The collision point is variable parameter.
For this study, most collisions are positioned at 1 au, with an eccentricity of zero putting the impact on a circular orbit. Though for a select few impacts, the semi-major axis is varied. The post-impact debris is given a bulk motion velocity, $v = \sqrt{GM_*/r}$, where $M_*$ is the mass of the central star and $r$ is the set semi-major axis, which would have been the orbital velocity of the progenitor planet prior to the impact. 
We note that a few impact velocities exceed or are comparable to the Keplerian velocity at 1 au which would be highly unlikely. However, these impacts are useful when considering extreme scenarios.

In the $N$-body code the condensate dust particles are treated like test particles and have no gravitational influence on each other. The largest remnant(s) gravitationally interact with each other and the dust. At the moment only the two largest post-collision remnants from the SPH simulations are used in the $N$-body code as most collisions will only produce one or two bodies that is massive enough to significantly influence the dust gravitationally. 

One problem we encounter in the hand off between the outcome from the SPH simulations and the input to the $N$-body simulations is the low particle number of vaporised escaping mass. We have solved this problem by upscaling the particle number. Because each particle in our SPH simulations contains a significant amount of mass, we can assume that each particle is an ensemble of particles that represents a size-distribution. We ensure that the spatial and velocity distributions are preserved in the following manner: we upscale every hand-off to $\sim$100,000 particles. 
Each unbound SPH particle, $P$, is used to generate $N$ particles randomly in a spherical volume defined by the four closest neighbours to that particle, where $N = 100000/P$ and $N$ is rounded to an integer value. Velocities are assigned using inverse distance weighting (see Appendix). This produces an upscaled distribution of particles that retains the original shape of the post-collision distribution. Another issue we encounter is that we do not know the actual orientation of the collision with respect to the orbit of the progenitor. To account for this we use two different orientations: 1) parallel to the progenitor orbit and; 2)  perpendicular to the progenitor orbit. These two orientations are the most extreme we can set, hence if orientation plays a role in disk evolution we will see the most change between these two orientations.

\subsection{Step 3: Synthetic Images (radiative transfer)}
To model the flux seen from a collision we use RADMC3D \citep{Dullemond2012-radmc3d}, a general-purpose package for modelling radiative transfer in three spatial dimensions using Monte-Carlo methods. The model requires inputs for dust density and structure, dust opacity and a source. The dust density and structure are determined from our $N$-body outputs, and the source is a Sun-like blackbody emitter of 5800 K. The dust opacities are determined from the opacity tool developed to determine the DIANA standard opacities \citep{Toon1981-Opacity,Woitke2015-opacity,Min2005-opacity,Dorschner1995-opacity}. The dust particles are dominated by silicates but with some carbonaceous material mixed in. Particles are given a size distribution of $dN/ds \propto s^{-3.5}$, with grain sizes raging from 0.5 $\rm{\mu m}$ to 1 mm. This size distribution is fixed at all time, i.e., no collisional evolution in the model. Particles are placed into $301^3$ bins with each side of the cubed grid set to 3 times the progenitor semi-major axis. Where over-densities occur, bins are sub-divided in order for a more realistic calculation. RADMC3D outputs images of spectral radiance at specified wavelengths which can then be converted to a flux from a system at a given distance away from the observer. We produce images at 24 microns. The spectral radiance is sampled every 0.1 orbits to produce a light curve. 

\section{Observability}
\label{section:observability}

Many factors will affect the observability of an extreme debris disk such as, the position of the collision around the central star, the parameters of the collision, and distribution of escaping material post-collision. The first factor we consider is the escaping vapour mass produced in embryo-embryo collisions and its lifetime.

\subsection{Vapour Production}

\begin{figure}
    \centering
    \includegraphics[width=\linewidth]{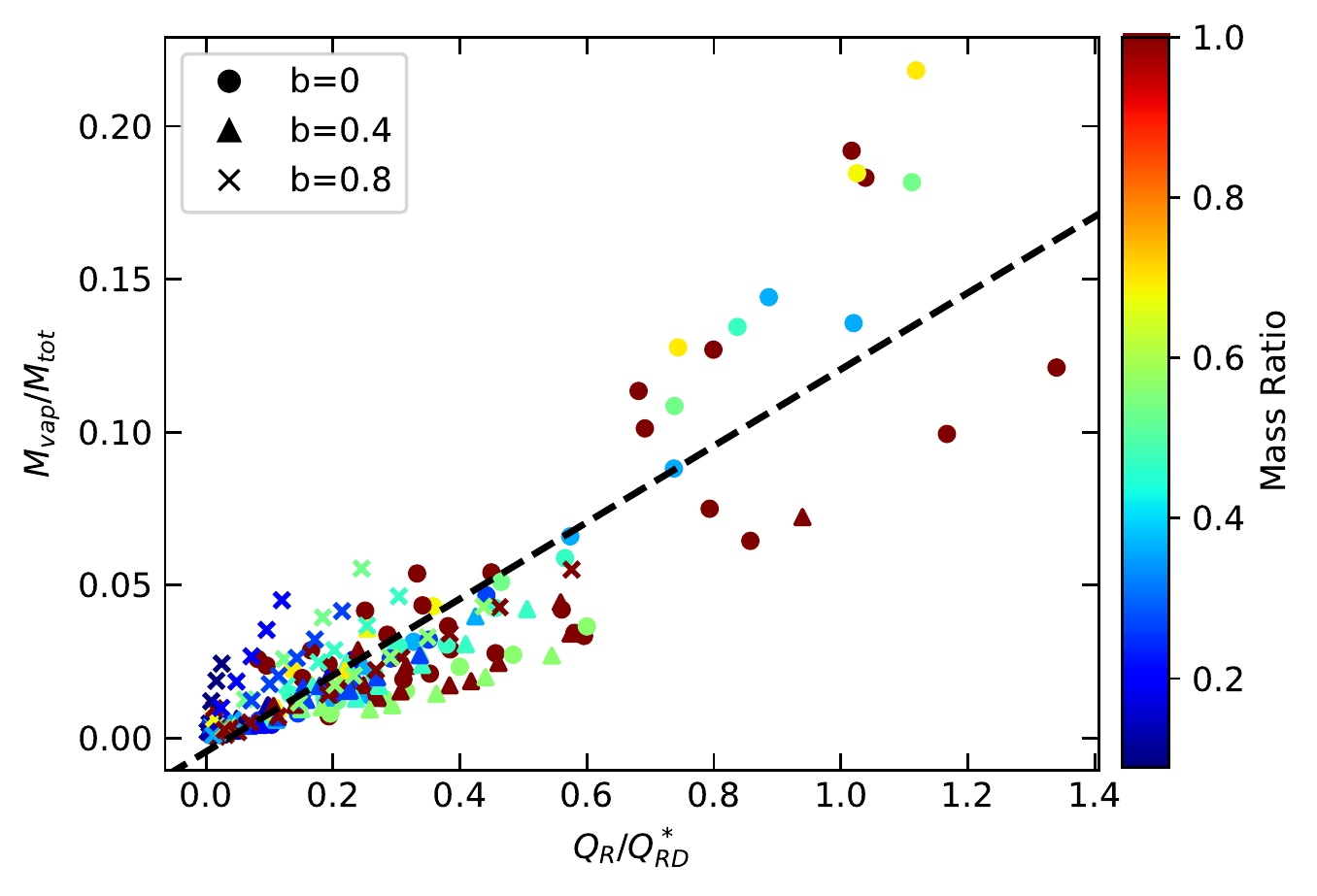}
    \caption{The mass of vapour ($M_{vap}$) in units of total mass ($M_{tot}$) produced in a given impact as a function of specific impact energy ($Q_R$) in units of catastrophic disruption threshold ($Q_{RD}^*$, the specific energy needed to permanently remove 50\% of the system mass). Normalising using $Q_{RD}^*$ allows us to know how destructive each collision was. Shape indicates impact parameter $b$, colour indicates mass ratio $\mu$.
    The dashed line shows the linear fit to the data.}
    \label{fig:mvapvseng}
\end{figure}

As stated in section \ref{section:methods}, we are focusing on the first observable debris disk produced by a giant impact created by the vapour condensate which forms small particles almost instantly post collision. To produce vapour, our collisions need to be super-sonic and shock-inducing which means impact speeds needed are  dependent on material type. For our modelled planetary embryos with the mantle made of forsterite, the impact speed needed is $\ge 6\ \rm{km\ s^{-1}}$. \citep{Davies_2019_vapour_speed,Davies_2020_vapour}. Collisions that are slower than this value will likely be fully dominated by melt material which form larger objects. Melt objects will collisionally evolve slowly as part of the background boulder population forming a traditional debris disk \citep{Johnson2014-meltdroplets}. We remove the melt mass as the focus of this paper is the early evolution of extreme debris disks which are dominated by vapour condensates formed directly from the giant impact.

Figure \ref{fig:mvapvseng} shows the vapour productions from all of the SPH impact simulations summarised in Table \ref{tab:collisions_less_1M_e} and \ref{tab:collisions_geq_1M_e} over a range of mass ratio and impact parameter. As expected we find that as impact energy is increased the fraction of vapour produced by the impact increases. To zeroth-order the relationship between specific impact energy ($Q_R$) and vapour mass appears linear and independent of collision parameters such as $\mu$ and $b$ with the data we have used. Using a least-squares fit to the data, we find the linear relation to be $M_{vap}/M_{tot} = 12.5\pm 0.5 \times 10^{-2}\,Q_R/Q_{RD}^* + 0.4\pm 0.2 \times 10^{-2}$ (black dashed line in Fig. \ref{fig:mvapvseng}), with errors indicating the level of scatter to 1 standard deviation in the data, where $Q_{RD}^*$ is the catastrophic disruption threshold which is the energy needed to disrupt half the mass in a collision. At high $b$ values there is some indication that the collision outcomes might deviate from the linear fit at $\gtrsim 1\ Q_R/Q_{RD}^*$, however such collisions would require an extreme impact velocity which would be unlikely in most orbital configurations. Further study is needed to determine the relationship between between low $\mu$, high $b$ collisions and escaping vapour mass produced.

The vast majority of escaping vapour in our simulated giant impacts is from mantle material. The linear fit shown in Fig. \ref{fig:mvapvseng} may break down at larger, more disruptive collisions when the core starts to make a significant contribution to the escaping vapour mass. This is because the position of the vapour dome for iron in temperature-specific energy space is different to that of forsterite. Iron has a larger triple-point temperature than forsterite, as a result, we might expect to see a discontinuity when a significant amount of the core vapourises. However, the simulations presented in this work extend from just above vapour production speed to 1.4 $Q_R/Q_{RD}^*$ \citep{Carter_2015_vel}, thus we cannot predict how the vapour production mass will scale with impact energy for very energetic giant impacts. 

\begin{figure}
    \centering
    \includegraphics[width=\linewidth]{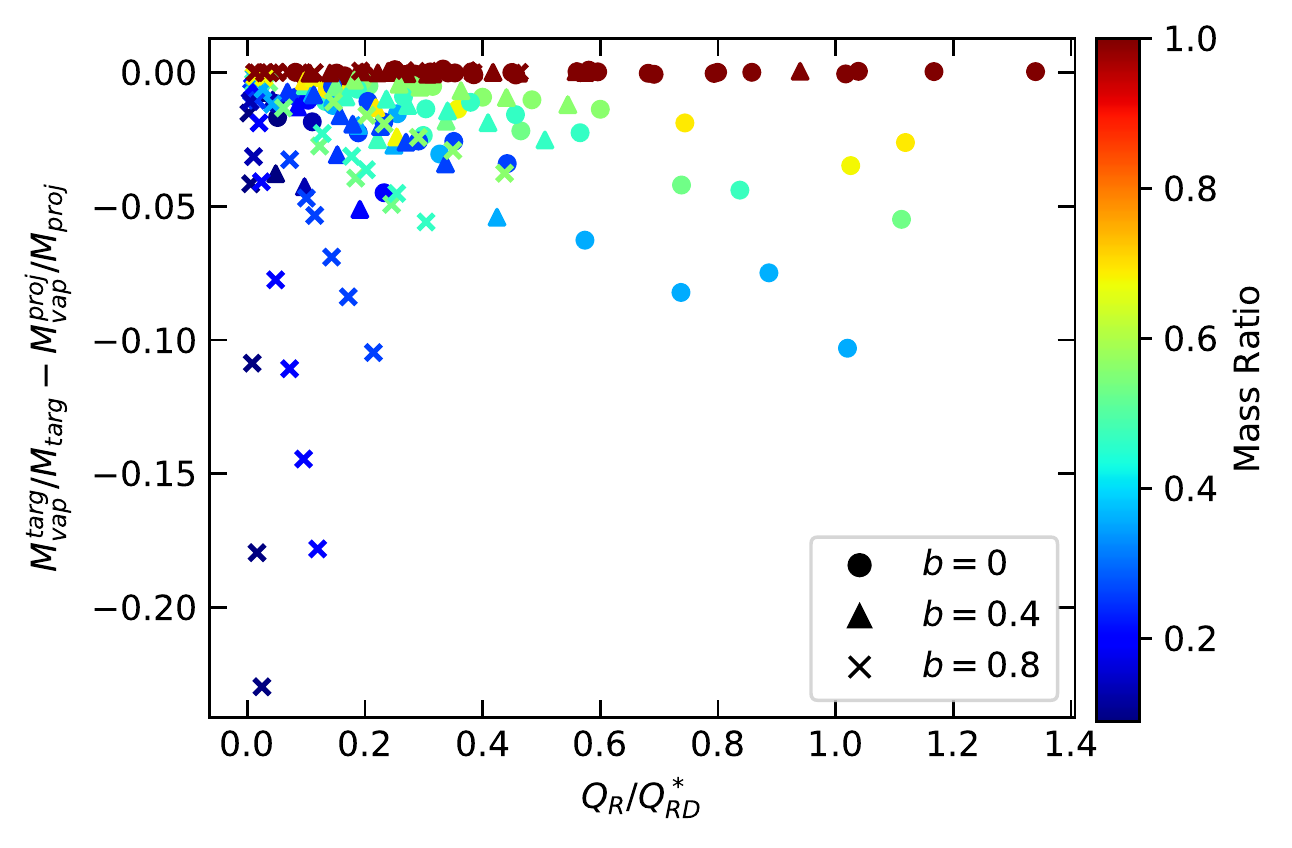}
    \caption{The difference between the escaping mass from the projectile normalised by projectile mass and the escaping vapour mass from the target normalised by the target mass versus the specific energy of the collision normalised by the catastrophic disruption criteria. Symbols and colour bar are same as in Fig. \ref{fig:mvapvseng}. 
    }
    \label{fig:M_vap_diff_eng}
\end{figure}

Although the amount of mass in vapour is relatively independent of $b$ and $\mu$ there is a strong dependence on these parameters when determining which body (target or projectile) the vapour comes from (Fig. \ref{fig:M_vap_diff_eng}). As expected for equal mass impacts, the vapour mass is produced equally from both the target and projectile, which is true for all impact angles (parameter $b$). However, as $\mu$ decreases, the fraction of vapour production from the projectile is higher than that of the target. This behavior is also sensitive to the impact angles as shown in Fig. \ref{fig:M_vap_diff_eng}.

\subsection{Initial Vapour Velocity Distribution}

\begin{figure*}
    \centering
    \includegraphics[width=\linewidth]{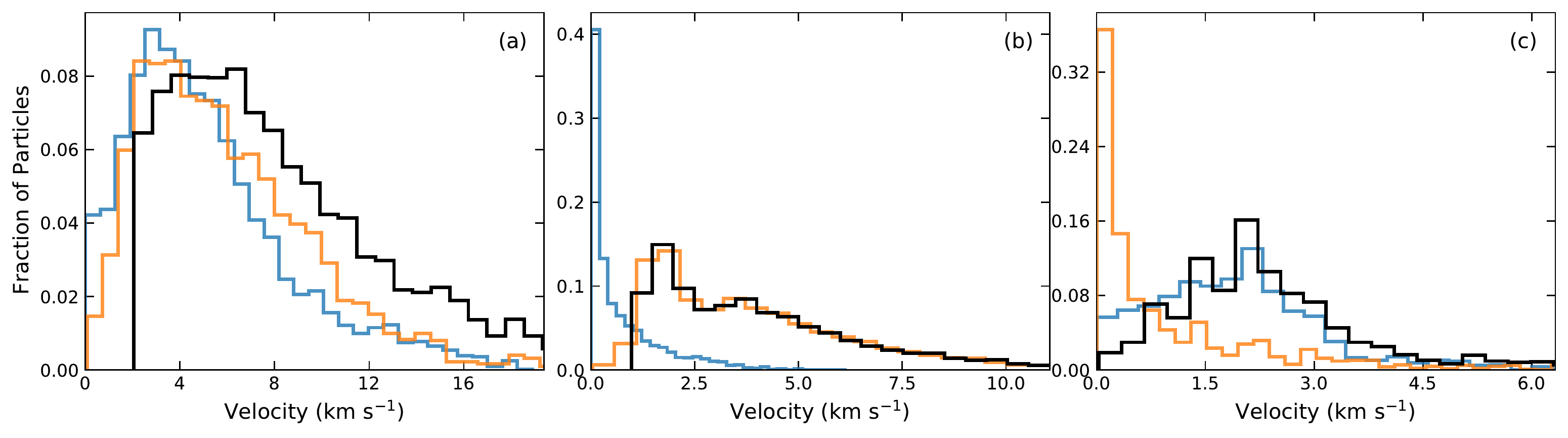}
    \caption{The absolute distribution of the launch velocities of vapour condensates with blue, orange, and black showing the parallel, perpendicular, and absolute velocity distributions respectively. (a) shows the distribution for sim 168, an impact between a 1.11 $\rm{M_{\oplus}}$ and 0.1 $\rm{M_{\oplus}}$ embryos with an impact velocity of 15 $\rm{km\ s^{-1}}$ at $b=0.4$, an impact condition similar to the canonical Moon forming impact. (b) shows the distribution for sim 8, an head-on impact between two 0.1 $\rm{M_{\oplus}}$ embryos with an impact velocity of 10 $\rm{km\ s^{-1}}$ and $b=0$. (c) shows the distribution for sim 88, impact between two 0.1 $\rm{M_{\oplus}}$ embryos with an impact velocity of 10 $\rm{km\ s^{-1}}$ and $b=0.8$.}
    \label{fig:vel_distrib_showcase}
\end{figure*}

Planetary embryo impacts produce a variety of post-collision dust distributions, none are isotropic as has been assumed in previous work \citep[e.g.][]{Jackson2012_Moon,Jackson-2014-planetary-collisions-at-large-au,Wyatt2017_dustsurvival}. To be able to compare different distributions, we need to characterise the anisotropic nature of the collisionally generated dust. We separate the velocity of each escaping particle into two velocity components, the velocity parallel ($v_{\parallel}$) and perpendicular ($v_{\bot}$) to the direction of the impact. For example, the direction of the impact is the same (along the x-axis) for the two collisions shown in Figure \ref{fig:embryocol}.
To understand and compare different distributions of particles, we choose to take the range between the 16\% and 84\% percentiles to compute the velocity dispersion for both velocity components in each collision ($\Delta v_{\parallel}$ and $\Delta v_{\bot}$). This allows us to compare the velocity distribution of different distributions without assuming a given distribution for any collision. We choose to take the 16\% and 84\% percentiles as this would be comparable to taking the 1$\sigma$ values either side of the mean of a normal distribution. 

In Fig.\ \ref{fig:vel_distrib_showcase} we show the absolute velocity distributions of vapour condensates parallel (blue) and perpendicular (orange) to the direction of the collision and absolute velocity (black) for three different giant impacts.
Fig.\ \ref{fig:vel_distrib_showcase}(a) shows sim 168, a collision between 1.11 $\rm{M_{\oplus}}$ and 0.1 $\rm{M_{\oplus}}$ at 15 $\rm{km\ s^{-1}}$ and $b=0.4$. We choose to show sim 168 as it is the closest simulated impact we have to the Moon-forming impact. Comparing to Fig.\ 1 of \citealt{Jackson2012_Moon}, we have an overall similar distribution of absolute velocities and we find only a small difference between the direction vapour condensates are launched (close to isotropic). Fig. \ref{fig:vel_distrib_showcase}(b) shows sim 8, a collision between two 0.1 $\rm{M_{\oplus}}$ at 10 $\rm{km\ s^{-1}}$ and $b=0$. Fig. \ref{fig:vel_distrib_showcase}(c) shows sim 88, a collision between two 0.1 $\rm{M_{\oplus}}$ at 10 $\rm{km\ s^{-1}}$ and $b=0.8$.
There is a stark difference between the direction vapour condensates are launched in sim 8 \& 88 compared to sim 168. By only using a truncated Guassian to fit to the absolute velocities, therefore assuming isotropic distribution, we would lose a lot of information of how vapour condensates are distributed in the highly anisotropic giant impacts.   
Due to the anisotropic nature of giant impacts, the orientation of the impact with respect to the orbit of the centre of mass has a surprising effect on the structure of the debris disk (see Section \ref{section:orientation}).

\begin{figure}
    \centering
    \includegraphics[width=\linewidth]{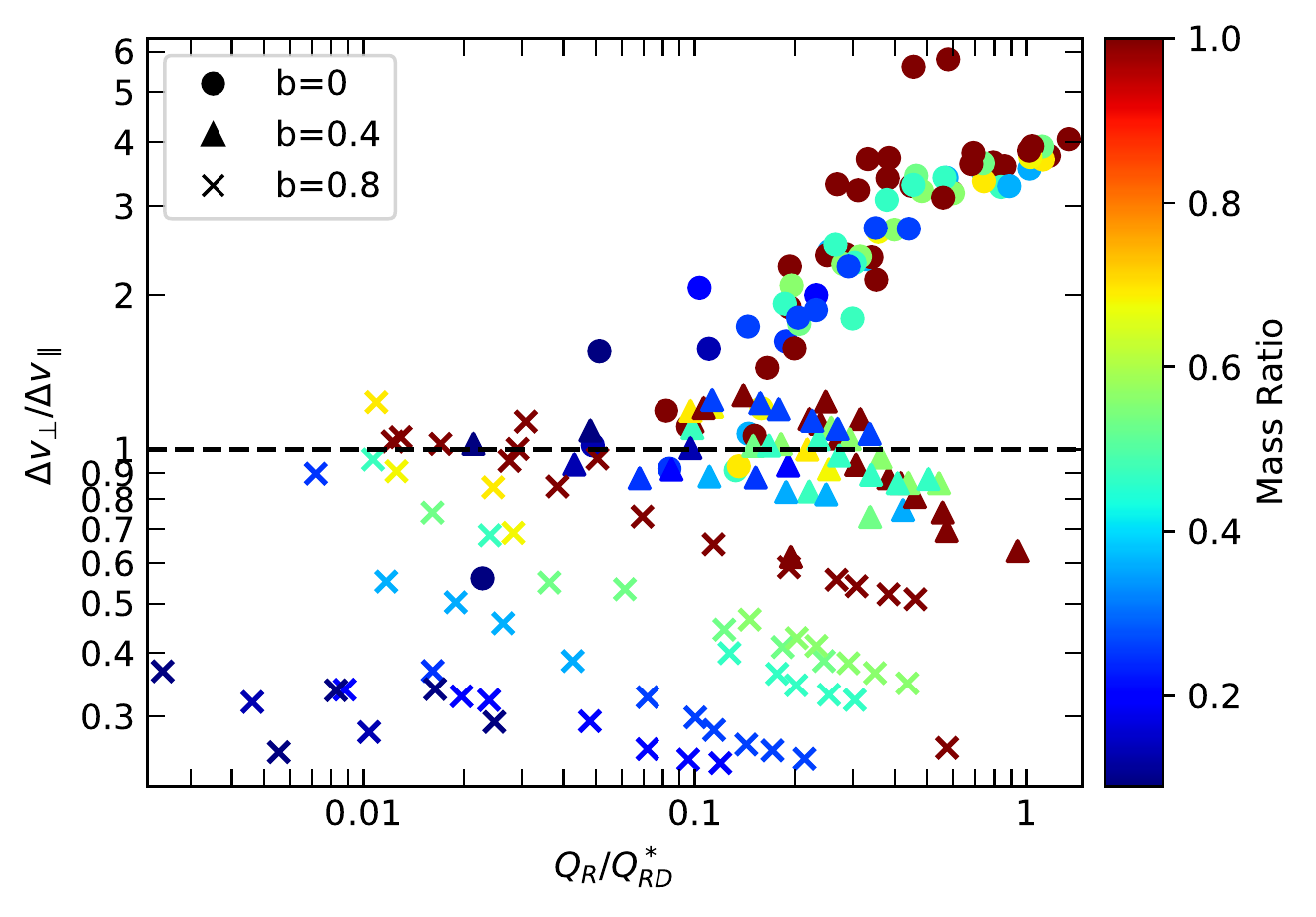}
    \caption{The ratio between the velocity dispersion perpendicular to the collision ($\Delta v_{\bot}$) and parallel to the collision ($\Delta v_{\parallel}$) versus the specific impact energy ($Q_R$) in units of catastrophic disruption threshold ($Q_{RD}^*$). Shape indicates impact parameter $b$, colour indicates mass ratio $\mu = M_{proj}/M_{targ}$. Dashed line shows velocity dispersion ratio of 1.}
    \label{fig:distrib_grid}
\end{figure}

\begin{figure}
    \centering
    \includegraphics[width=\linewidth]{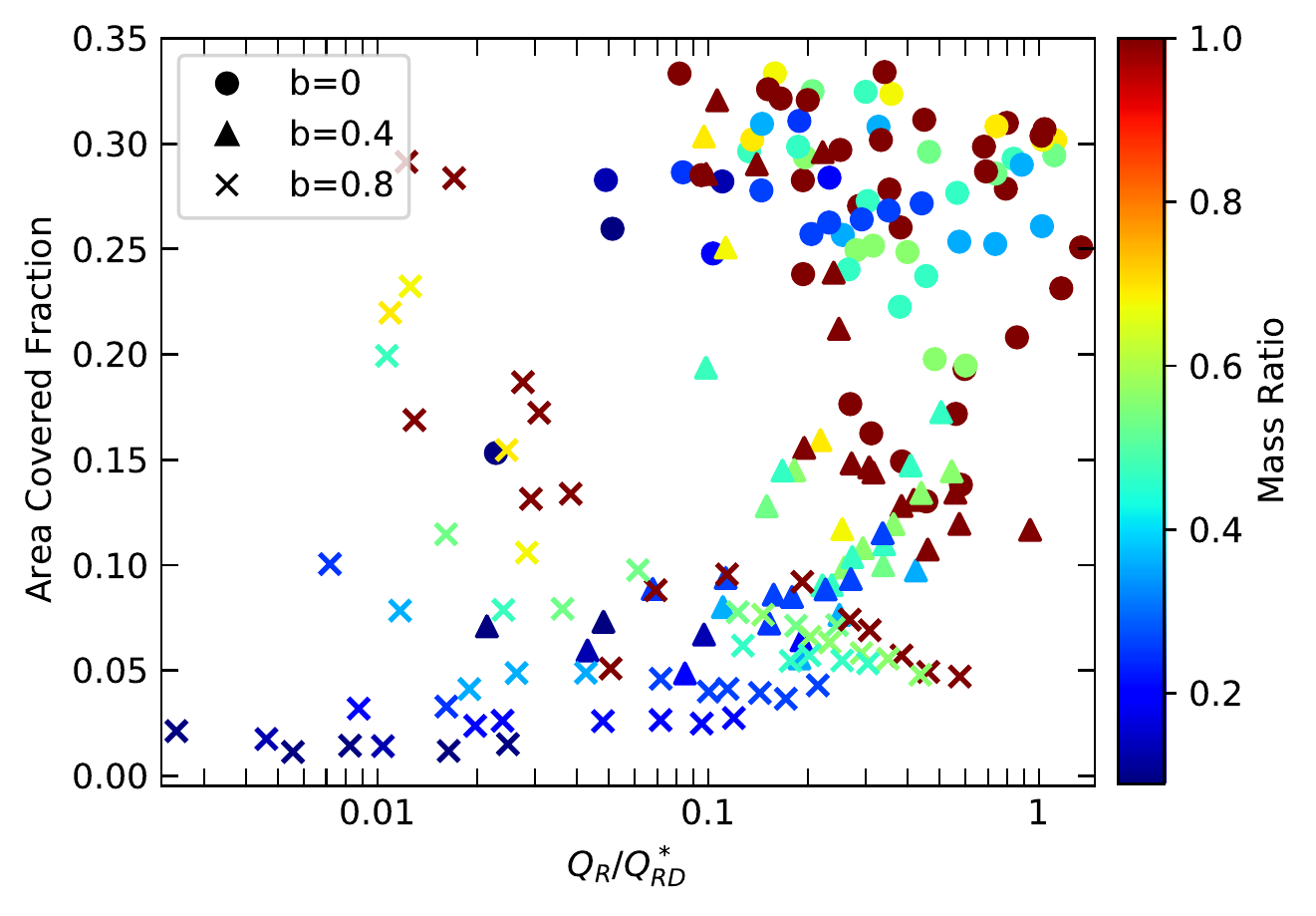}
    \caption{Anisotropy of collisions. The surface area covered by the particle velocity vectors projected onto the surface of a sphere for each collision. An isotropic distribution will have a value close to 1, a very anisotropic distribution will have a value close to 0. Shape indicates impact parameter $b$, colour indicates mass ratio $\mu = M_{proj}/M_{targ}$.}
    \label{fig:anisotropy}
\end{figure}

Figure \ref{fig:distrib_grid} shows the ratio between the velocity dispersion perpendicular and parallel to each impact against the specific energy of the impact normalised by the catastrophic disruption threshold. We can see the impacts are separated into three distinct groups defined by their impact parameter. 
Figure \ref{fig:anisotropy} shows the surface area covered by the particle velocity vectors when projected onto a sphere versus specific energy of the impact normalised by the catastrophic disruption threshold.  Isotropic distributions will have a value close to 1, while a value close to 0 will mean a very anisotropic distribution. Collisions will likely kick particles in all directions, but a low area covered fraction will indicate clumping of velocity vectors in one or more directions.
This difference in distribution of velocity will have an effect on the initial visibility of the dust which we will discuss in section \ref{section:orientation}.

For head-on impacts, $b=0$, the majority of the impact debris is distributed toward the direction that is perpendicular to the impact direction, i.e., $\Delta v_{\bot}$ dominates for most $Q_R/Q_{RD}^*$ values. Mass ratio does not influence $\Delta v_{\bot}/\Delta v_{\parallel}$, and the ratio is a function of specific impact energy as the higher energy impact gives a larger ratio . Near 0.8 $Q_R/Q_{RD}^*$ the increase in $v_{\bot}/\Delta v_{\parallel}$ becomes shallower and potentially may plateau; however this needs to be confirmed by future simulations with more impacts at larger $Q_R/Q_{RD}^*$. The increase in $\Delta v_{\bot}/\Delta v_{\parallel}$ with $Q_R/Q_{RD}^*$ can be explained by looking at strip (a) in Fig.\ref{fig:embryocol}. Impacts colliding head-on "pancake", pushing material perpendicular to the direction of the collision. Some material will be released parallel when material collapses back down, re-shocking the material. 
The area covering fraction of head-on impacts in Fig. \ref{fig:anisotropy} mostly fall between 0.25 and 0.35. This is expected as material is preferentially kicked perpendicular to the collision, but we see no variation across different $Q_R/Q_{RD}^*$ or mass ratio values. So for larger $Q_R/Q_{RD}^*$, the ratio of debris kicked perendicular to parallel stays the same but debris kicked perpendicular will have a wider velocity distribution.

For impacts at $b=0.4$, $\Delta v_{\bot}/\Delta v_{\parallel}$ stays around a value of 1 with a slight decrease at larger $Q_R/Q_{RD}^*$. Similarly, there is no variation with mass ratio. 
This suggests the $b = 0.4$  collisions might be isotropic, but Fig. \ref{fig:distrib_grid} is misleading in isolation. If we consider the results shown in Fig. \ref{fig:anisotropy} we see that the $b = 0.4$ collisions are not isotropic as most of these impacts lie between 0.05 and 0.15 area covering fraction indicating that these collisions are very anisotropic, more so than the head-on impacts. There also seems to be a  relationship between mass ratio and area covered, with lower mass ratios typically having a lower area covering fraction. So for these impacts to have a value roughly close to $\Delta v_{\bot}/\Delta v_{\parallel}=1$ means debris must be preferentially launched close to $45^{\circ}$ to the impact. 
At larger $Q_R/Q_{RD}^*$, we see a decline in $\Delta v_{\bot}/\Delta v_{\parallel}$ which might be caused by the projectile not fully interacting with the target, meaning a proportion of material is ejected in the direction of the collision. 
There also might be a mass ratio dependence. In Fig. \ref{fig:anisotropy}, the area covering fraction seems to decrease with larger $Q_R/Q_{RD}^*$ for impacts with mass ratios close to 1, while mass ratios close to 0.1 show no variation. This suggests that impacts with mass ratios close to 1 become more anisotropic with increasing $Q_R/Q_{RD}^*$. One explanation for this is these impacts change in how debris is launched. Though more impacts above 1 $Q_R/Q_{RD}^*$ are needed at different mass ratios to test this.

High impact angle, grazing  impacts ($b=0.8$) show behaviour which varies with the mass ratio of the impact. For large mass ratios (close to 1), the interacting mass fraction between the target and the projectile means both will survive the collision creating two remnants of similar masses, with some mass ejected parallel to the collision, giving small $\Delta v_{\bot}/\Delta v_{\parallel}$. The area covering fraction of most $b=0.8$ impacts is small compared to $b=0$ and $b=0.4$ impacts, further suggesting the distribution is jet-like. At lower mass ratios, the projectile is preferentially disrupted over the target leading to material being ejected in the direction the projectile was moving in. This is backed up by the smaller area covering fractions for low mass ratio impacts. At low $Q_R/Q_{RD}^*$, $b=0.8$ impacts seem to have a large area covered for larger mass ratio impacts. These impacts may not follow the jet-like distribution we would expect from a high impact angle collision. While this is interesting, this might be caused by small number statistics because only a small number of particles escape in these low energy, high impact angle collisions. 

\subsection{Orientation}

As we have discussed, post-collision velocity distributions are anisotropic and this has a profound effect on the early architecture of the extreme debris disk in determining both the spatial and flux evolution of the dust. 

\label{section:orientation}
\subsubsection{Spatial Evolution}
\begin{figure*}
    \centering
    \includegraphics[width=\linewidth]{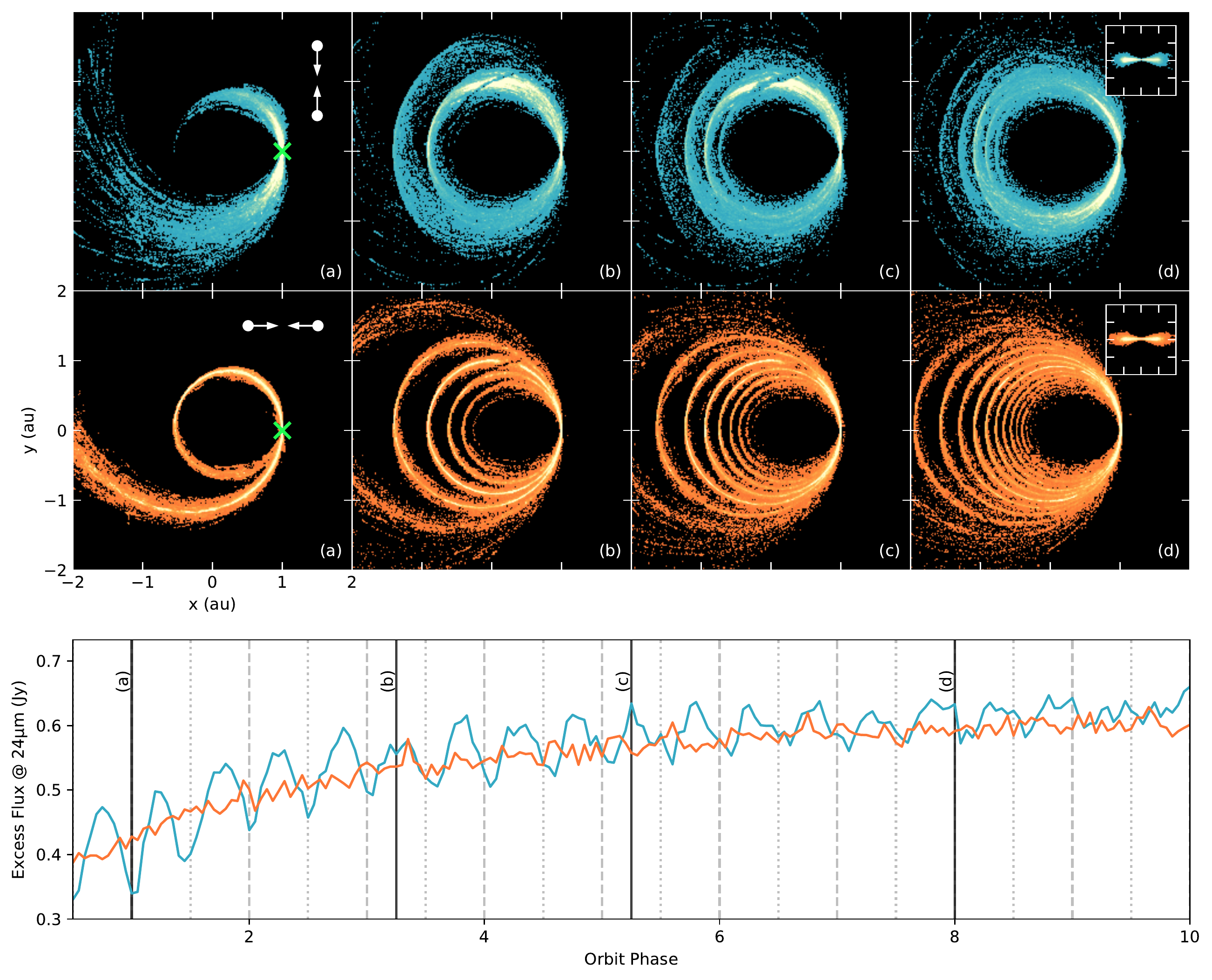}
    \caption{The time evolution of the dust clump released at 1 au from a head-on collision between two $0.1\ \rm{M_{\oplus}}$ embryos at $10\ \rm{km\ s^{-1}}$. The top panel shows four snapshots of the post-collision dust for two different orientations rotated by 90$^{\circ}$ as illustrated by the white dots with arrows signifying the orientation of the collision with respect to the progenitor orbit at the launch position located in the (a) panels. The green cross denotes the collision point. The bottom panel shows the flux evolution at $24\ \rm{\mu m}$ as the clump evolves assuming a face-on view of each disk with the line colours matching the snapshot colours above. Snapshots represent the number density of particles with the maximum showing 30 particles to show the structure of the disk. The snapshot times are shown in the flux plot by black solid lines with notations which relate back to the correct snapshot below.
     The greyed dashed and dotted lines show the expected orbit phase position of dips in flux occurring from the collision point and anti-collision line respectively. The inserts in the (d) panels show the edge-on (y--z) view of the disk with the collision point in the centre.
     }
    \label{fig:flux_and_nbody}
\end{figure*}

\begin{figure*}
    \centering
    \includegraphics[width=\linewidth]{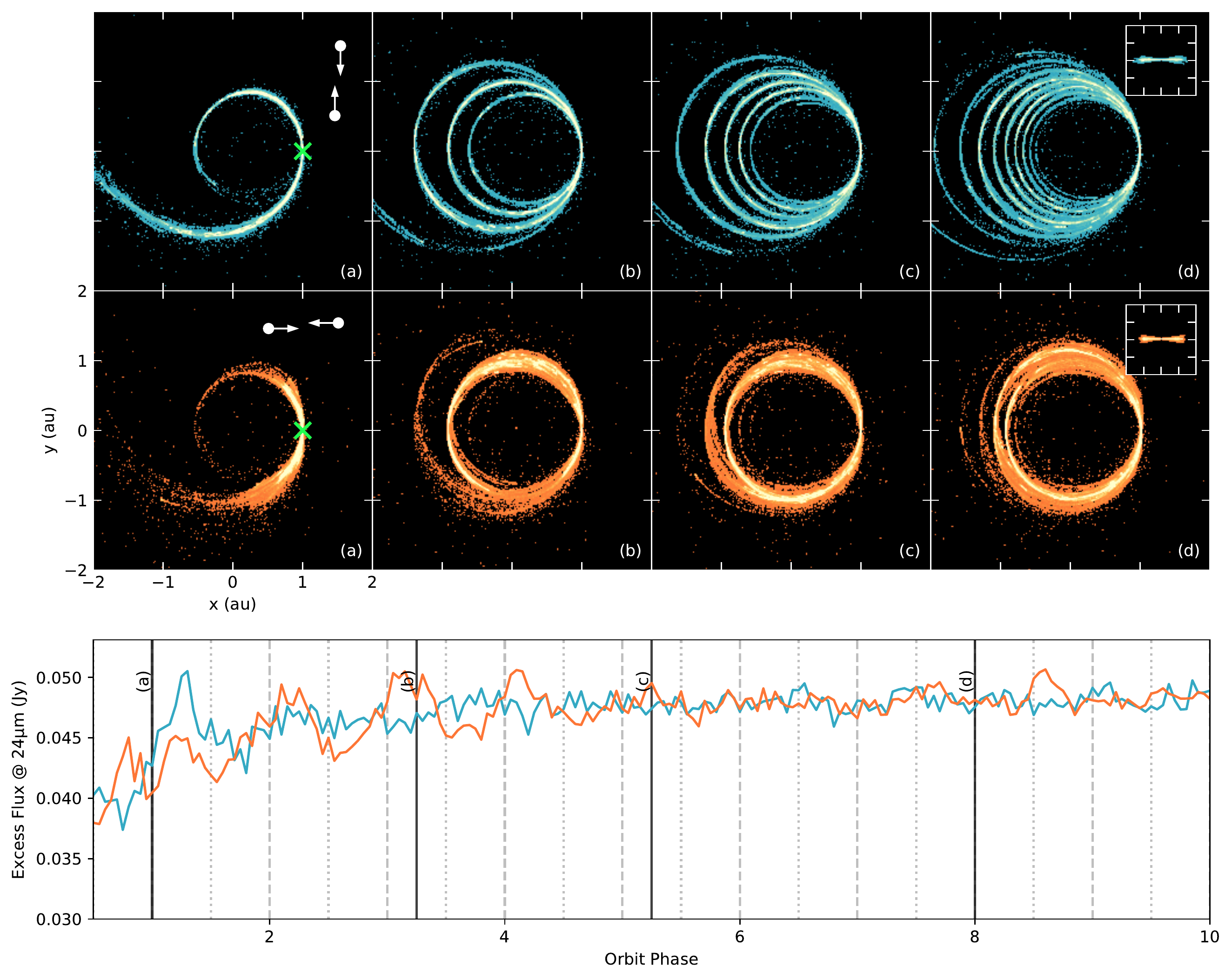}
    \caption{Same as in Fig. \ref{fig:flux_and_nbody} but now with sim 88 in table \ref{tab:collisions_less_1M_e} which is a collision between two $0.1 \rm{M_{\oplus}}$ embryos at 10 $\rm{km\ s^{-1}}$ with an impact parameter of 0.8.}
    \label{fig:flux_nbody_hr}
\end{figure*}

Due to the anisotropy of the dust post-collision, a collision between two embryos with the same collision parameters but different orientations with respect to the progenitor orbit can produce debris disks which do not look alike. Figures \ref{fig:flux_and_nbody} and \ref{fig:flux_nbody_hr} show 
 time series of the vapour generated dust (increasing in time from left to right) from a head-on, equal-mass impact (sim 8) and a grazing equal-mass impact (sim 88) for two different collision orientations. The orientation of the impact is indicated in the upper right of frame (a). The top row in blue shows the dust distribution for an impact that occurs along the orbit whereas the orange time series in the second row shows the distribution resulting from an impact perpendicular to the orbit.
 
We find the overall disk evolution to be similar to that found in \citealt{Jackson2012_Moon} and \citealt{Jackson-2014-planetary-collisions-at-large-au}, with the initial dust clump quickly transitioning into a spiral arm feature within 1 to 2 orbital phases. The spiral arm winds up over the subsequent orbital phases, forming concentric rings. However, there are some interesting differences in the details of the evolution of the four simulations. In Fig. \ref{fig:flux_and_nbody}, the orange disk, where the impact is perpendicular to the progenitor orbit, transitions from the spiral arm phase into the concentric ring phase quicker than the corresponding blue disk, where the impact was oriented along the orbit. The rings in the orange disk are more frequent and expand over a wider range of semi-major axis.

We see similar behaviour for sim 88 shown in Fig. \ref{fig:flux_nbody_hr}. Equal mass, hit-and-run collisions preferentially launch vapour condensates along the direction of the collision rather than perpendicular like equal mass, head-on collisions (Fig. \ref{fig:embryocol} and Fig. \ref{fig:distrib_grid}). Fig. \ref{fig:flux_nbody_hr} also shows four snapshots between the same two objects at the same speed as in Fig. \ref{fig:flux_and_nbody} but now at $b=0.8$ (sim 88) instead of $b=0$ (sim 8). We see that sim 88 shows similar structures to that of sim 8, though for opposite orientations (blue in Fig. \ref{fig:flux_and_nbody} is similar to orange in Fig. \ref{fig:flux_nbody_hr} and vice versa). For sim 88, the distribution of vapour parallel to the remnants post-collision mean the velocity kicks will mostly align with the initial progenitor velocity vector with the opposite orientation to sim 8. This is why we see similar structures but for opposite orientations for the simulated disks shown in Fig. \ref{fig:flux_and_nbody} and Fig. \ref{fig:flux_nbody_hr}.

\begin{figure*}
    \centering
    \includegraphics[width=\linewidth]{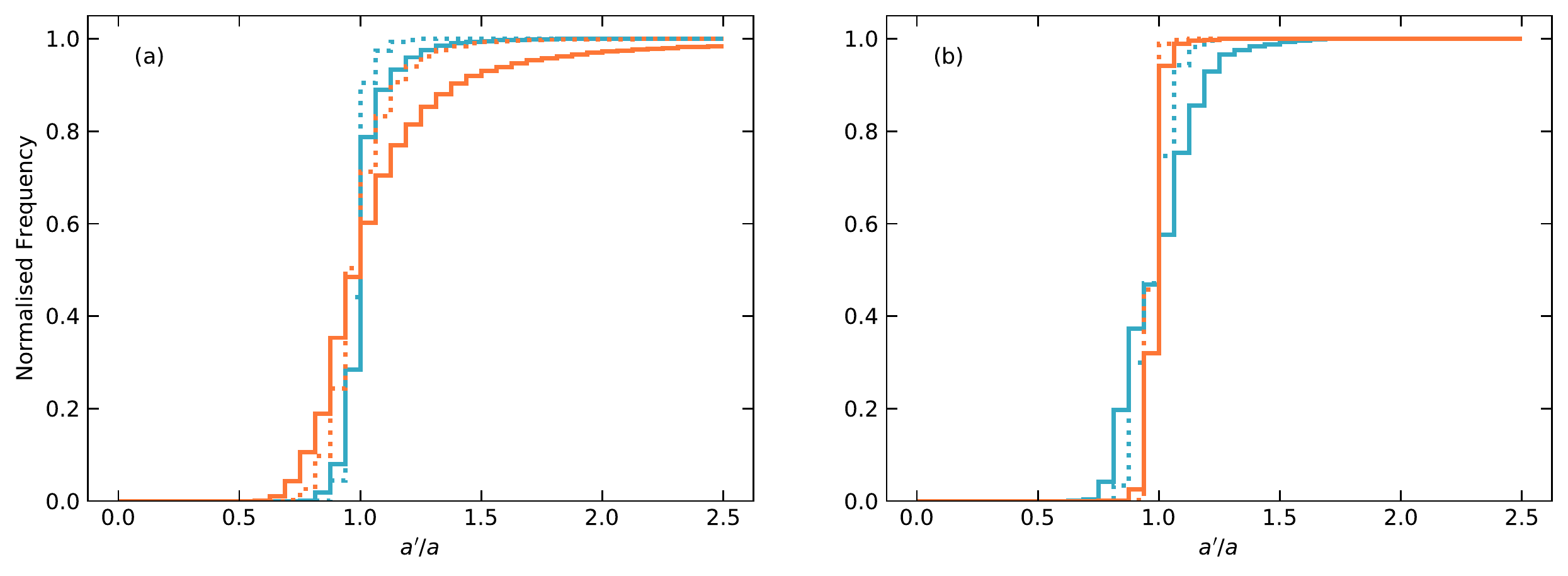}
    \caption{Cumulative histograms showing the distribution of semi-major axis values for dust post-collision for a collision between two $0.1\ \rm{M_{\oplus}}$ embryos at $10\ \rm{km\ s^{-1}}$ at (a) $b=0$ (sim 8) and (b) $b=0.8$ (sim 88), normalised by the semi-major axis of the progenitor. The colour denotes the rotation of the collision, same as in Fig. \ref{fig:flux_and_nbody} and \ref{fig:flux_nbody_hr}. Collisions at 1 au are solid lines while dotted lines are collisions at 0.3 au. The orientation will change the distribution of orbital parameters. For a given collision, distance from the star will also dictate the distribution of dust, with larger range of semi-major axis values available at larger distances. }
    \label{fig:sma_cumul_hist}
\end{figure*}

Figure \ref{fig:sma_cumul_hist} shows the initial semi-major axis distributions for the different orientations in both Fig. \ref{fig:flux_and_nbody} and Fig. \ref{fig:flux_nbody_hr}. The semi-major axis of particles are determined at $t=0$ in the $N$-body code, and are derived from adding the velocity kick (determined from the pass over of data between the SPH simulations and $N$-body code) to the circular orbit of the progenitor. A clear difference is observed in the distributions of vapour condensates between the different orientations, and is present in both examples of sim 8 and 88. When we look at Fig. \ref{fig:embryocol} and Fig. \ref{fig:distrib_grid}, we can see that the vapour is not distributed isotropically for sim 8 and 88, but the vapour. is ejected in a preferred direction(s). The distribution of velocity kicks cause this anisotropic distribution of vapour. Therefore, the orientation of the collision will determine the direction(s) of where the majority of vapour will be given velocity kicks. The direction and magnitude of a velocity kick will determine how much the new orbit of a vapour condensate will differ from the circular orbit of the progenitor orbit. For a given magnitude of velocity kick, the largest change in semi-major axis happens
when the direction of the velocity kick is parallel with the initial velocity and when the initial velocity is at the maximum, hence at the perigee (any position in on a circular orbit). \citealt{Jackson-2014-planetary-collisions-at-large-au} goes into detail in how orientation of the velocity kick will influence the resulting orbit. For sim 8, the orientation shown in the orange disk of Fig. \ref{fig:flux_and_nbody} means vapour condensates are preferentially launched along a direction that is parallel to the velocity of the progenitor. Vapour condensates are launched perpendicular to the progenitor velocity in the blue disk. 

\subsubsection{Flux Evolution}
The flux evolution of the disks is shown below the snapshots in both Fig \ref{fig:flux_and_nbody} and \ref{fig:flux_nbody_hr}. The colour of the flux evolution matches the snapshot colour (blue for the impact oriented along the orbit and orange for the impact oriented perpendicular to the orbit). At $24\rm{\mu m}$, we find a similar overall flux for the different orientations of the same collision. The disks generated from different orientations of the same collision contain the same vapour condensate mass, therefore similar overall flux levels are to be expected.

For the head-on equal-mass impact orientated parallel to the orbit (blue line) shown in Fig. \ref{fig:flux_and_nbody}, we find short term variation (wiggles) up to $\sim 10$ orbits after the collision. This short-term variation is similar to that seen in extreme debris disks like ID8 and P1121 \citep{Meng2015-pc,Su-2019-extreme-disk-variability}. However, the short-term variation does not appear when the impact is rotated perpendicular to the orbit (orange line). Again the difference in behaviour between the two collision orientations is due to the difference in spatial distribution. The distribution of orbital parameters (semi-major axis, eccentricity, inclination, etc) for vapour condensates is smaller in the blue disk compared to the orange disk. In the blue disk, the distribution of orbital parameters leads to the disk being optically thick at the collision point and anti-collision line for many orbits, reducing the flux when the progenitor was expected to pass through these points. For the orange disk, the distribution of orbital parameters means the disk shears out quickly and the disk is not optically thick at the collision point/anti-collision line after 1-2 orbits. 

In comparison we see no sustained short-term variation in the flux evolution of the equal-mass grazing collision shown in Fig. \ref{fig:flux_nbody_hr}. If we only consider the spatial distribution, we would expect to see the orange disk shown in Fig. \ref{fig:flux_nbody_hr} to behave similarly to the blue disk shown in Fig. \ref{fig:flux_and_nbody} but this is not the case. The difference is the total mass ejected as vapour. For sim 88 shown in Fig. \ref{fig:flux_nbody_hr}, the escaping vapour mass was $3.79\times 10^{-4} \rm{M_{\oplus}}$. For sim 8 shown in Fig. \ref{fig:flux_and_nbody}, the escaping vapour mass was $6.64\times 10^{-3} \rm{M_{\oplus}}$. Optical thickness of the disk is dependent on how dense the disk is. The reduced mass released from sim 88 due to the projectile only partially colliding with the target means that at 1 au, the resulting disk becomes optically thin quickly after the collision has taken place and a maximum flux ($\sim$10 times less) is reached by two orbital phases.

\subsubsection{Producing Wiggles}

\begin{figure}
    \centering
   \includegraphics[width=\linewidth]{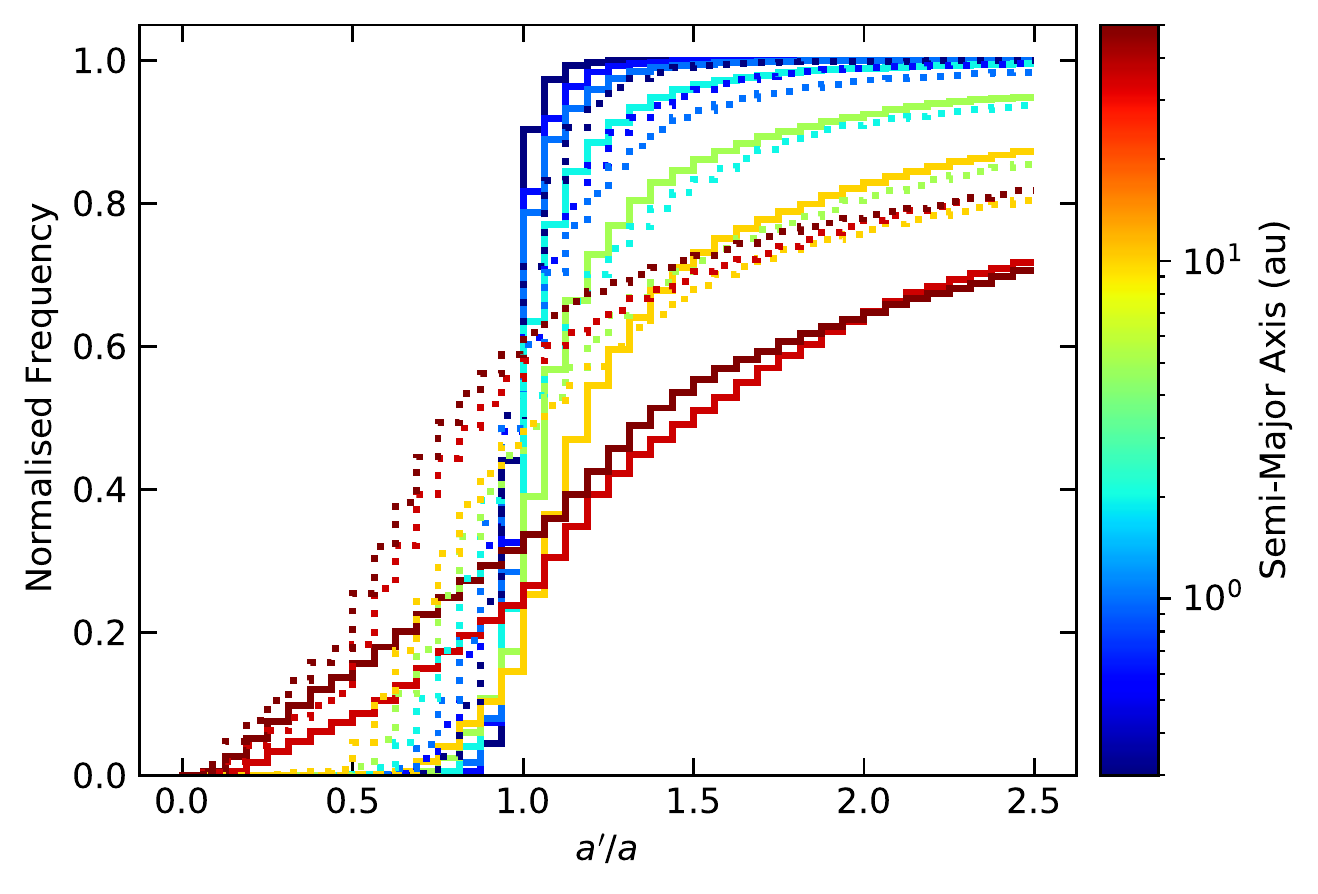}
    \caption{Cumulative histograms of the semi-major axis distributions of the initial vapour condensates from a head-on collision between two $0.1\ \rm{M_{\oplus}}$ embryos at $10\ \rm{km\ s^{-1}}$ (sim 8) normalised by the semi-major axis of the progenitor. The colour of the line denotes the semi-major axis of the progenitor. Solid dotted lines show the collision orientated parallel and perpendicular with the velocity vector of the progenitor respectively.}
    \label{fig:sma_hist_au}
\end{figure}

It is possible for sim 88 to show short-term variation if placed closer to the star because this will lead to a reduction in the distribution of the orbital parameters (Fig. \ref{fig:sma_cumul_hist} and \ref{fig:sma_hist_au}) and therefore increase the optical thickness of the disk, causing the debris at the collision point and anti-collision line to be optically thick for more orbits. 
However, if this collision occurred much closer to the star, the relative difference between the optical thickness of material passing through the collision point/anti-collision line is reduced and wiggles will not appear. Though this is for face-on disks, for inclined disks wiggles can still arise from material passing through the disk ansae.
There will be a defined range of orbital parameters and orientations for each collision where the short-term variation in the flux evolution can occur.

There are a few other ways to increase the optical thickness of the disk so that a grazing impact like sim 88 could produce sustained wiggles in the orange flux shown in Fig. \ref{fig:flux_nbody_hr}  similar to the blue flux in Fig. \ref{fig:flux_and_nbody}: 1) reduction in the impact velocity which will reduce the range of velocity kicks; 2) or an increase in total mass involved in the collision. 
We have to be mindful though that escaping vapour mass and the velocity dispersion of the escaping vapour are not independent from each other. Increasing the impact velocity will increase the escaping vapour mass but will also increase the velocity dispersion. Because high $b$ collisions are inefficient producers of escaping vapour mass, to produce the oscillations in the flux similar to that seen in Fig. \ref{fig:flux_and_nbody}, the distance at which the collision can take place from the star is more limited than head-on collisions.

Head-on collisions are the most efficient producers of escaping vapour, therefore, can have a wider range of impact velocities and hence distances from the star which the short-term variation in the flux evolution can occur. One way to increase the escaping vapour mass but keep the impact velocity the same is to increase the mass involved in the collision. But increasing the mass involved in the collision increases the mutual escape speed of the collision. The impact velocity needs to be greater than the mutual escape speed in order to eject vapour condensates and place the dust on orbits that are not bound to the collision remnants. 

This will also limit the distance from the star at which giant impacts can occur at and reproduce the short-term variation seen in the flux from the resulting disk. A coherent clump of material will not be formed if a giant impact takes place where the mutual escape speed is greater than or equal to the Keplerian velocity at that position as material ejected will be placed onto extremely wide distributions of eccentricity and semi-major axis values \citep{Wyatt2017_dustsurvival}, with a portion of the material ejected from the stellar system.

Collisions closer to the star will be more likely to produce the short-term variation in the flux evolution.
Nevertheless, the evolution of impact-produced debris closer to the star will evolve much quicker than a disk further out. Therefore, it becomes increasingly difficult to observe such oscillation behaviour if the impact-produced debris disperses after a few tens of orbits.  
There is a sweet spot between all the parameters we have discussed, with the observability of oscillatory behaviour arising in disks formed from giant impacts decreasing as we move away from this sweet spot. 

\section{Discussion}

In this paper, we have demonstrated how short term variations ("wiggles") in the flux evolution can be expressed or suppressed in an extreme debris disk formed from a specific giant impact. 
\subsection{Progenitor Position}

For a given collision, the radial position of the progenitor orbit will also affect the orbits of the escaping material and thus the production or suppression of "wiggles". For example, a collision that takes place further from the host star will have a wider range of parameter space to fill because the effect of the gravitational potential well is reduced (Fig. \ref{fig:sma_hist_au}). 
In other words, we find that impacts that occur close to the host star result in the escaping dust being placed onto a narrow range of semi-major axis about the progenitor's semi-major axis. While impacts that occur at larger stellocentric distances have escaping debris in a wider range of semi-major axis. (Note that Fig. \ref{fig:sma_hist_au} was constructed under the same impact velocity at different stellocentric distances; therefore the result does not take into account the likelihood of such an impact at that location due to its orbital Keplerian velocity).
The widening of the semi-major axis distribution as the collision moves further away from the star leads to the reduction and eventual disappearance of any "wiggles" in the flux evolution.  
This is because that the "wiggles" rely on the vapour condensates being a coherent clump as it passes through the collision point and anti-collision line, meaning that oscillations only happen when the there is a narrow distribution of dust in the semi-major axis. 

If the distribution of eccentricities in the impact-produced debris is too wide, and therefore the semi-major axis range is large, the dust will Keplerian shear out on a timescale before oscillations can be observed. At large distances from the host star, we find the impact orientation becomes more influential because the semi-major axis distribution is broader. This is because the velocity kick of the vapour condensates only changes direction while the Keplerian orbital velocity becomes smaller. If the material is optically thin, the broader semi-major axis distributions would then display a variety of dust temperatures. A change in orientation the could lead to a change in the dust temperatures as seen, for example, by the red solid and dashed lines in Fig. \ref{fig:sma_hist_au}. As the observed flux is sensitive to the dust temperature, the same collision but with different orientations could look very different when observed. For collisions occurring at large semi-major axis, this effect would be very pronounced. 

\subsection{Orientation Probability}

Wiggle behaviour is also sensitive to the orientation of the impact relative to the progenitor orbit. The effect of varying the orientation is dependent on the initial vapour velocity distribution of the impact. We have shown in Fig. \ref{fig:distrib_grid} how the initial vapour velocity distribution for many impacts varies with impact parameter and mass ratio. 
For an impact which creates enough escaping vapour and occurs close enough to the star so the Keplerian speed is large compared to the velocity dispersion (a coherent clump can form), and all orientations are equally likely then there will be an orientation for a given impact that will produce "wiggles".
However, if the orientation probability distribution is skewed (for example, impacts parallel to the progenitor orbit are more likely than perpendicular to the progenitor orbit for some dynamical reason) then depending on what orientations are favoured, wiggles will be expressed or suppressed (Fig. \ref{fig:flux_and_nbody} and \ref{fig:flux_nbody_hr}).
Oblique impacts are more likely to occur than head-on collisions \citep{shoemaker_1962}, hence if the orientation distribution is skewed to favour forming "wiggles" from ejecta launched perpendicular to a collision (head-on collisions) then "wiggles" are less likely to occur.  

\subsection{Dust Survivability}

The observability of an extreme debris disk is dependent on both how bright it is and how long the dust within the system survives. Up to this point we have focused solely on the initial disk made from vapour condensate and have ignored dust generated by the intermediate boulder population, but in order to discuss lifetime we must consider all of the mass available to generate observable dust. As a reminder to the reader we are considering a single energetic giant impact between two planetary embryos. This impact has occurred as a result of planet formation and is a growth generating impact. For simplicity let us consider a head-on impact like simulation 8. This impact results in one significant remnant, some vapour ejected in a disk perpendicular to the impact, and non-vapour debris and melt. It is this non-vapour ejecta that we will call the boulder population. Initially, the vapour, the boulder population, and the largest post-collision remnant are all on similar orbits, however, the vapour and the boulders do have velocity perturbations as a result of the impact that place them on slightly different orbits. 

In this work we assume that the vapour condenses into a small characteristic size immediately. Because of the small condensate size the vapour population is observationally visible immediately with very little, if any, collisional evolution. This means that the dust generated by the vapour becomes visible to the observer while it is still on similar orbits to the largest remnant - the vapour generated dust appears while it is still in a "clump". The boulder population, however, is most likely not visible immediately. Although we do not have the resolution in our SPH simulations to confidently determine the full size distribution of the non-vapour post-collision remnants we do have the resolution to identify the second (and third) largest remnants. In simulation 8 the second largest remnant is about 100 km and previous work on asteroid families indicates a size distribution with a 3.5 index power-law. This means that the majority of the mass will be in the larger end of the power-law. Assuming that a traditional collisional cascade is initiated by orbit crossing between the boulders it will take some time for enough collisions to occur to generate an observable dust population. Thus, we assume that by the time the orbits of the boulder population have spread out and started crossing they will no longer be in an identifiable clump. Using this logic we have come up with a simplified model to estimate the lifetime and observability of debris produced by the giant impact modeled in simulation 8.

We can make an estimate of how long we would expect escaping material from an isolated giant impact to last by assuming the escaping debris forms a debris disk fed by a quasi-steady state collisional cascade. 
Using equation 15 from \citealt{Wyatt-2008-evolution-of-debris-disks}, 
\begin{equation}\label{eqn:wyatt08}
    f/M_{tot}=0.37r^{-2}D_{bl}^{-0.5}D_c^{-0.5},
\end{equation}
we can find the fractional luminosity $f$ (the luminosity of the debris disk divided by the luminosity of the host star), where $M_{tot}$ is the total escaping mass, $r$ is the position of the disk in au, $D_c$ is the size of the largest object in km, $D_{bl}$ is the blowout size in $\rm{\mu m}$.
In following this method we have made the following simplifications/assumptions: 1) the debris disk is an axisymmetric narrow ring; 2) there is an underlying population with the largest size bin set at $D_c$; 3) the disk follows the typical size distribution of a normal debris disk/collisional cascade ($dN\propto a^{-3.5}$), and 4) material is instantly lost once below the blow-out size of the star and no longer contributes to the flux. 

If we assume $f = 0.01$ is required to produce an observable debris disk and calculate $M_{tot}$ for a traditional debris disk taking reasonable values of 60 km and 0.8 $\mu m$ for $D_c$ and $D_{bl}$ respectively, then at 1 au, around a solar-type star, equation \ref{eqn:wyatt08} gives us a mass of $1.87\times10^{-1}\rm{M_{\oplus}}$. This result means that a giant impact would need to release at least a Mars mass amount of material into the surrounding environment to form a detectable extreme debris disk. Assuming a standard collisional cascade, the size of the largest boulders sets the lifetime of the debris disk, the larger the largest objects the longer the disk will last. 
 We have shown in Fig. \ref{fig:mvapvseng} that giant impacts between planetary embryos will generally lose between a few to 10\% of their mass as escaping vapour.
 If we instead assume that the escaping vapour mass is solely responsible for the debris disk (assuming no underlying km-scale boulders) and thus fix $D_c = 100\rm{\mu m}$ then the total mass to create $f =0.01$ at 1 au becomes $7.64\times 10^{-6}\ \rm{M_{\oplus}}$. If we assume a more conservative size distribution for the condensed dust between mm-cm the mass needed might be as large as $2.42\times 10^{-3}\ \rm{M_{\oplus}}$. 
 However, the lifetime of a disk populated by small vapour condensates would be much less than a traditional debris disk  because there is no significant reservoir with which to resupply the dust. So although an extreme debris disk may be created almost instantaneously from a giant impact it may be fleeting and thus difficult to detected.

\begin{figure*}
    \centering
    \includegraphics[width=\linewidth]{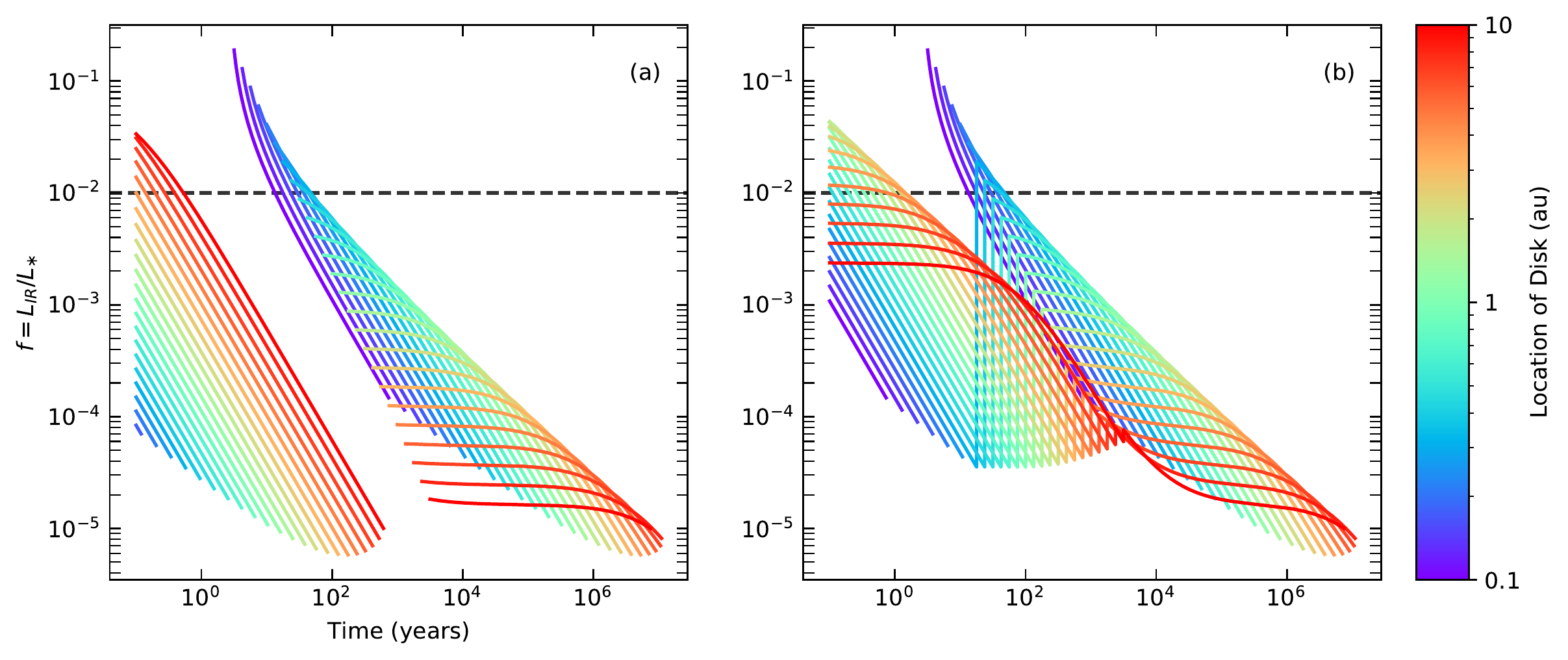}
    \caption{Evolution of fractional luminosities of compound debris disks with maximum initial grain sizes of: (a) $100\rm{\mu m}$, and (b) $100\rm{mm}$ at t=0. Each sub-figure shows the time evolution for disk locations varying from 0.1 au to 10 au. Lines are plotted up to the detection limit at $24\rm{\mu m}$ (comparison of wavelength dependent limits shown in Fig. \ref{fig:frac_lims}). The mass of each disk is determined from a head-on collision between two $0.1\rm{M_{\oplus}}$ embryos at $10\rm{km\ s^{-1}}$. Initially the debris disk is assumed to be formed entirely from vapour condensate. After 100 orbits dust contributions from the rest of the unbound post collision mass is added to the flux. The black dashed line, $f=0.01$, is the minimum fractional luminosity limit for extreme debris disks.}
    \label{fig:f_vs_time_dmax}
\end{figure*}

\begin{figure}
    \centering
    \includegraphics[width=\linewidth]{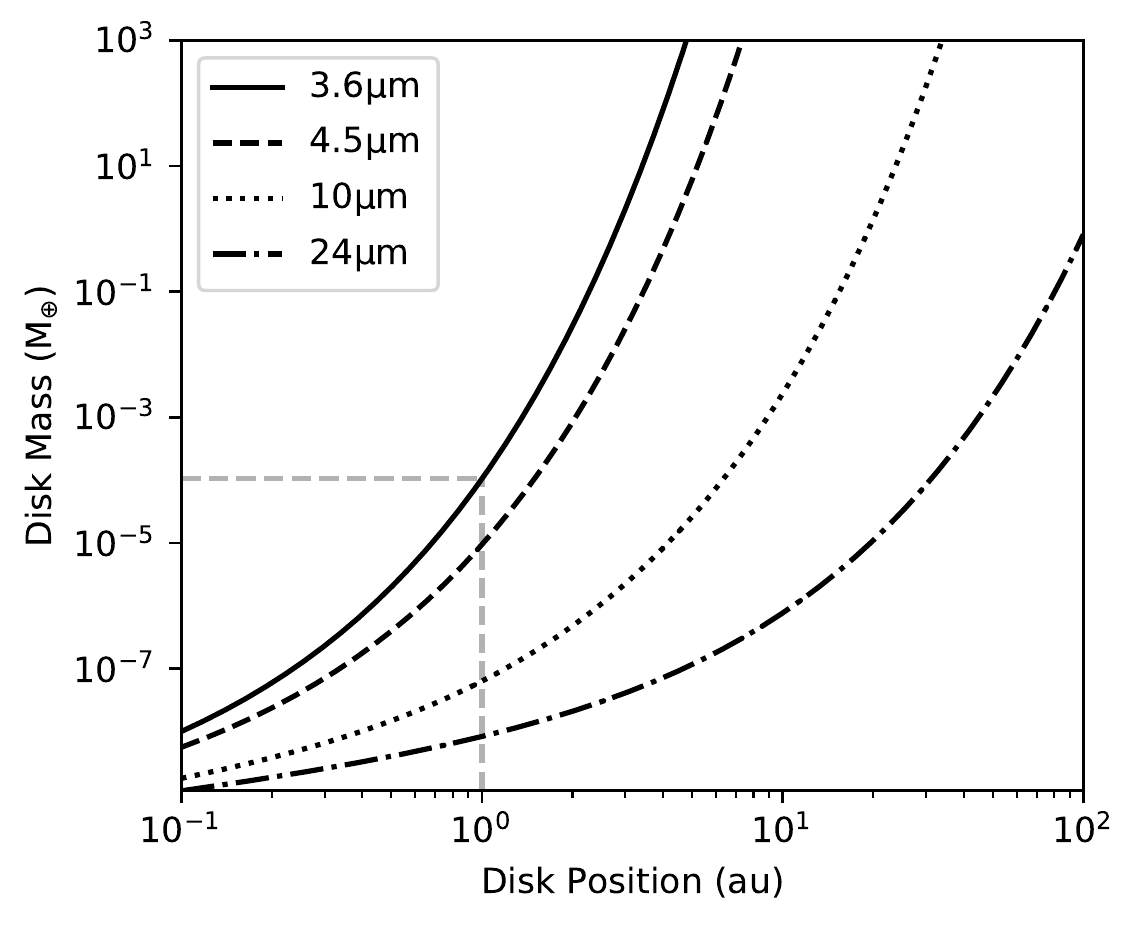}
    \caption{The minimum disk mass needed to observe a debris disk around a Solar-like star at varying disk positions. The lines correspond to different wavelengths of calibration limited fractional luminosity calculated assuming an excess ratio, $R_v\ = \ F_{v \rm{disk}}/F_{v *}$, of 0.03 \citep{Wyatt-2008-evolution-of-debris-disks}. Disk masses assume a maximum size of 100$\rm{\mu m}$ in the disk. The gray dashed lines intersect the $3.6\rm{\mu m}$ line to show the minimum mass needed to observe a debris disk at 1 au. For Spitzer missions which observe at $3.6\rm{\mu m}$ and $4.5\rm{\mu m}$, dust created from embryo collisions is unlikely to be observed past 3 au as a substantial amount of mass will be needed.}
    \label{fig:frac_lims}
\end{figure}

As mentioned in the first paragraph we would expect a combination of a vapour condensate debris disk and a traditional debris disk formed from the grinding of the km-scale boulders produced in the impact. We would expect the vapour condensate debris disk to be brightest almost immediately after the impact because the maximum condensate size is small while we would expect the boulder debris disk to take many orbits to evolve into a steady state collisional cascade. This idea is expressed in Fig. \ref{fig:f_vs_time_dmax} which shows how the fractional luminosity varies with time for a compound debris disk produced by sim 8 (Table \ref{tab:collisions_less_1M_e}) for two values of the initial vapour condensate $D_c$, $100\ \rm{\mu m}$ and $100\ \rm{mm}$, and maximum boulder size of 100km for a range of radial locations from 0.1 au to 10 au. In this model we assume that the visible vapour condensate debris disk is formed immediately while the traditional boulder generated debris disk takes 100 dynamical times/orbits to develop and create a quasi-steady state collisional cascade. In addition, we are assuming for simplicity, that the two dust populations do not interact with each other.

To construct figure \ref{fig:f_vs_time_dmax} we begin with the total escaping mass from the result of the SPH simulation. In sim 8, the escaping vapour mass is $6.64\times 10^{-3}\rm{M_{\oplus}}$ and the escaping non-vapour/boulder mass is $4.5\times 10^{-2}\rm{M_{\oplus}}$ (Tab. \ref{tab:collisions_less_1M_e}). 
 Each curve is calculated by using eqn. \ref{eqn:wyatt08}, with the mass varying with time as:
\begin{equation}
    M_{\rm{tot}}(t) = M_{\rm{tot}}(0)/[1+(t-t_{stir})/t_c]
\end{equation}
from \citealt{Wyatt-2008-evolution-of-debris-disks}, where,
\begin{equation}
    t_c = 1.4\times 10^{-9}r^{13/3}(dr/r)D_cQ_D^{*5/6}e^{-5/3}M_*^{-4/3}M_{\rm{tot}}^{-1},
\end{equation}
$dr/r$ is the width of the disk which is set to 0.5, $Q^*_D$ is the planetesimal strength assumed to be 150 $\rm{J\ kg^{-1}}$, $e$ is the mean planetesimal eccentricity determined from the eccentricity of the vapour condensates, and $M_*$, the central star mass, is one solar mass. The timescale of mass loss ($t_c$) starts when the destructive collisions occur which is determined by $t_{stir}$. For the vapour generated disks we assume destructive collisions between vapour condensates occur immediately $t_{stir} = 0$, while the boulder population has $t_{stir} = 100$ dynamical times. For $t < 100$ dynamical times we assume that the boulder population does not contribute significantly to the flux. The initial mass ($M_{\rm{tot}}(0)$) is the total vapour condensate/boulder mass bound to the star, any vapour condensate/boulder particle unbound ($e>1$) is removed from the total mass.

Each curve is plotted until the fractional luminosity falls below the calibration limit needed to be observed at $24\rm{\mu m}$ for a given semi-major axis (Fig. \ref{fig:frac_lims}).
The calibration limit is set so that the excess flux, $R_v = F_{v\rm{disk}}/F_{v*}$ where $F_{v\rm{disk}}$ and $F_{v*}$ are the flux from the debris disk and star respectively, has to be above 0.03. Using equation 11 from \citealt{Wyatt-2008-evolution-of-debris-disks} we can determine the fractional luminosity detection limit, $f_{\rm{det}}$ for a debris disk at a given wavelength, $\lambda$, via,
\begin{equation}
    f_{\rm{det}} = 6\times 10^9 R_v r^{-2} L_* T_*^{-4} B_v(\lambda ,T_*)[B_v(\lambda ,T)]^{-1}X_{\lambda},
\end{equation}
where $r$ is the distance to the disk from the star, $L_*$ is the luminosity of the star, $T_*$ is the blackbody temperature of the star, $T$ is the temperature of the disk, $B_v$ is the blackbody emission, and $X_{\lambda}$ is a factor included to take account of the falloff in the emission spectrum at large wavelengths but it's 1 for $24\rm{\mu m}$.
Figure \ref{fig:frac_lims} shows how $f_{\rm{det}}$ varies with disk mass and position for four different wavelengths: 3.5, 4.5, 10, and 24$\rm{\mu m}$. As radial distance increases the disk mass needed for detection increases as does the amount of escaping material that is unbound. This creates the positive inflection in the limiting $f_{det}$ in fig. \ref{fig:f_vs_time_dmax}. 

Now if we consider the results shown in fig. \ref{fig:f_vs_time_dmax} starting first with a modest head-on a giant impact (sim 8) close to the parent star (blue and purple curves) we see that it results in an initially brief and faint vapour condensate disk followed by a break and a bright but quickly fading traditionally formed debris disk. In this case only the second boulder generated debris disk would be briefly considered an extreme debris disk ($f>0.01$). If instead the collision occurred further from the central star at 10 au the situation is effectively reversed (red curves). The vapour generated disk is initially bright and in the case of 100 $\mu$m observationally classed as an extreme debris disk for a few years but fades three orders of magnitude in flux over the next 100 dynamical times or so until the slowly evolving boulder population produces a faint traditional debris disk that lasts for millions of years. Impacts that occur in the terrestrial region result in an intermediate outcome, 
resulting in initial disks that only last a few years at most. The boulder disks will appear afterwards. Increasing the initial condensate size from 100 $\mu$m to 100 mm results in vapour disks that are observable for longer (due to a modest reservoir) with more vapour disks being classed initially as extreme debris disks. In addition, the disconnect between the vapour and boulder disks is removed except disks placed closer than $\sim$0.4 au to the star. 

Note the results presented in fig. \ref{fig:f_vs_time_dmax} do not allow any interactions between the two disks though the total flux from each of the vapour produced disk and the traditional boulder debris disk are included, however, the dust from the vapour disk does not interact with the boulder population dynamically. In addition, the debris disks produced from sim 8 do not stay bright for extended periods of time and although this is primarily due to the small amount of escaping mass available for the debris disk the simplifying assumptions discussed at the beginning of this section will also lead to the most efficient removal of dust.  Namely, the dust is assumed to be in a fully formed axisymmetric disk, with an evenly distributed removal of material around the disk. But we know that this is not entirely accurate. 
After a giant impact, the initial escaping debris clumps will be asymmetric with collisions between debris more likely at the collision point and anti-collision line \citep{Jackson-2014-planetary-collisions-at-large-au,Kral-2017_asymmetry_disk}. This asymmetry will make the evolution of the disk significantly faster than a axisymmetric disk. Therefore, we would expect to the lifetime of the boulder population to be shorter than that seen in Fig. \ref{fig:f_vs_time_dmax}. However, we expect it still takes many interactions for the boulder population to grind down and produce a detectable amount of small grains, thus, we should still see a period of time between the vapour disk declining in flux and the boulder disk then increasing in flux.
The asymmetry will also affect the vapour disk but because we have used a strength value, $Q^*_D$, an order of 1-2 magnitudes lower than what is expected for small grains \citep{B&A_1999_small_grain_strength}, the lifetime of the vapour disks should largely stay the same. 

In this model, the vapour disk has a lifetime which is not consistent with observed extreme debris disks. ID8 has multiple instances where the disk shows extreme fractional luminosity with disk positions less than 1 au with timescales longer than what we find in Fig. \ref{fig:f_vs_time_dmax}. We know this mass cannot be sustained by a large reservoir of boulders due to the sharp decline in flux in 2013 and 2015 following a large increase in flux in 2012 and 2014 \citep{Su-2019-extreme-disk-variability}. Therefore, there must be a mechanism which sustains the small vapour grains which we are not accounting for. One way to allow the vapour disk to survive for longer is to have a mechanism which will protect grains at, and smaller than, the blow-out size. Future work is needed to understand what this mechanism could be. 

In presenting this idea of a compound debris disk we are making the assumption that the two components of the flux excess are expected to behave differently dynamically. The traditional boulder debris disk forms from an azimuthally distributed collisional cascade not by a single large event that results in some orbital coherence of the debris at particular locations in the disk. Thus, we would expect any observed short term variations like the "wiggles" seen in extreme debris disks ID8 and P1121 are most likely produced by the vapour condensate disk only. However, it is possible that under some particularly active scenarios the boulder population may also be able to produce short term variations.

In this section we have only presented the flux versus time of one collision (sim 8), however, Table \ref{tab:collisions_less_1M_e} and \ref{tab:collisions_geq_1M_e} show the collision outcomes are diverse. An increase or decrease in the debris mass,
 which we have found to vary between $1\times10^{-4}$ to one Earth mass in our simulated collisions, will lead to an decrease/increase in the initial fractional luminosity but overall the evolution of the disks will be similar. Decreasing the mass (vapour/boulder or both) will lead to some disks at large orbital distances in Fig. \ref{fig:f_vs_time_dmax} becoming unobservable as there is not enough mass in the disk. Meaning that some of these giant impacts would not be observable at 24 $\mu m$ but it would not mean that the giant impacts did not occur. Note that disks at large orbital distances are observed at longer wavelengths. This changes the observability threshold and might mean disks that are not detectable at 24 $\mu m$ are detectable at longer wavelengths. 

\section{Conclusions}

This study focused on the early behaviour of extreme debris disks formed from giant impacts between planetary embryos. The goal was to numerically model the formation of a vapour condensate debris disk along with any short-term variations or "wiggles" which could be linked to observations of variable extreme debris disks such as ID8 and P1121. We constructed a hybridised numerical model using SPH (modified version Gadget 2) to calculate the energetic impact between planetary embryos and determine the distribution of impact induced vapour. We then simulated the global evolution of the vapour condensed dust using an $N$--body code. 

From the giant impact simulations we determined the dependence of vapour mass on impact energy creating a first-order scaling law that can be used in future work instead of the computationally costly numerical impact simulations. We also showed that a greater percentage of vapour mass is ejected from the projectile than the target as you move towards lower mass ratios and larger impact parameter values, with the difference becoming more profound at larger $Q_R/Q_{RD}^*$. We then showed how the material is preferentially launched from an impact and how that varies with mass ratio and impact parameter. Material is more likely to be launched perpendicular to the impact direction at larger $Q_R/Q_{RD}^*$ if the impact is head-on ($b=0$) with no variation with mass ratio, while for grazing impacts with large impact parameter ($b=0.8$) material is launched parallel to the impact direction, with the preference to launch parallel becoming larger as the mass ratio becomes smaller. For intermediate impact parameter values ($b=0.4$) material is launched in all directions.

Orientation of a giant impact and the subsequent vapour distribution with respect to the progenitor orbit plays a large role in whether short-term variation is seen in extreme debris disks or not. 
We find that we can remove the short-term variation seen on one of our extreme debris disks through orientating the impact differently so that the vapour condensates have a wider distribution of eccentricites and semi-major axis values. This leads to material shearing out more quickly, meaning there is no coherent clump of material that is optically thick at the collision point and anti-collision line after 1-2 orbits.

Finally, we discussed the lifetime of giant-impact induced disks. We modeled giant impact induced disk as a compound disk made of two distinct populations, vapour condensate and dust created by grinding boulders produced in the original impact. The evolution was calculated through semi-analytical means following \citealt{Wyatt-2008-evolution-of-debris-disks}.
In our simple evolution model we assumed that the vapour condensate disk was produced immediately and that the traditional boulder debris disk took 100 dynamical times to develop. With these parameters we found that vapour production alone would not guarantee a classification as an extreme debris disk. In fact depending on the location of the impact, the disk formed from the grinding of the boulder population was the only component that would be observationally characterised as an extreme debris disk. In addition, it was clear that in most cases the disks were only visible for very short periods of time, making observational detection difficult and strongly dependent on the nature of the impact and the amount of vapour and boulder mass produced. 

This work suggests that giant impacts produce a complex compound debris disk that is in most cases variable and transient. The current small numbers of detections of young extreme debris disks do not mean that giant impacts are not occurring but that we have been incorrect about their observability.

\section*{Acknowledgements}

 LW acknowledges financial support from STFC (grant S100048-102). KS is grateful for funding from NASA's ADAPs programs (NNX17AF03G and 80NSSC20K1002). This work was carried out using the computational facilities of the Advanced Computing Research Centre, University of Bristol - http://www.bris.ac.uk/acrc/. Thanks to Phil Carter, Jack Dobinson and anonymous reviewer for useful discussion that improved the quality of this project. 

\section*{Data Availability}
The data underlying this article are available in the article.




\bibliographystyle{mnras}
\bibliography{main.bib} 
\appendix
\section{}
\subsection{Inverse Distance Weighting}
\begin{figure}
    \centering
    \includegraphics[width=\linewidth]{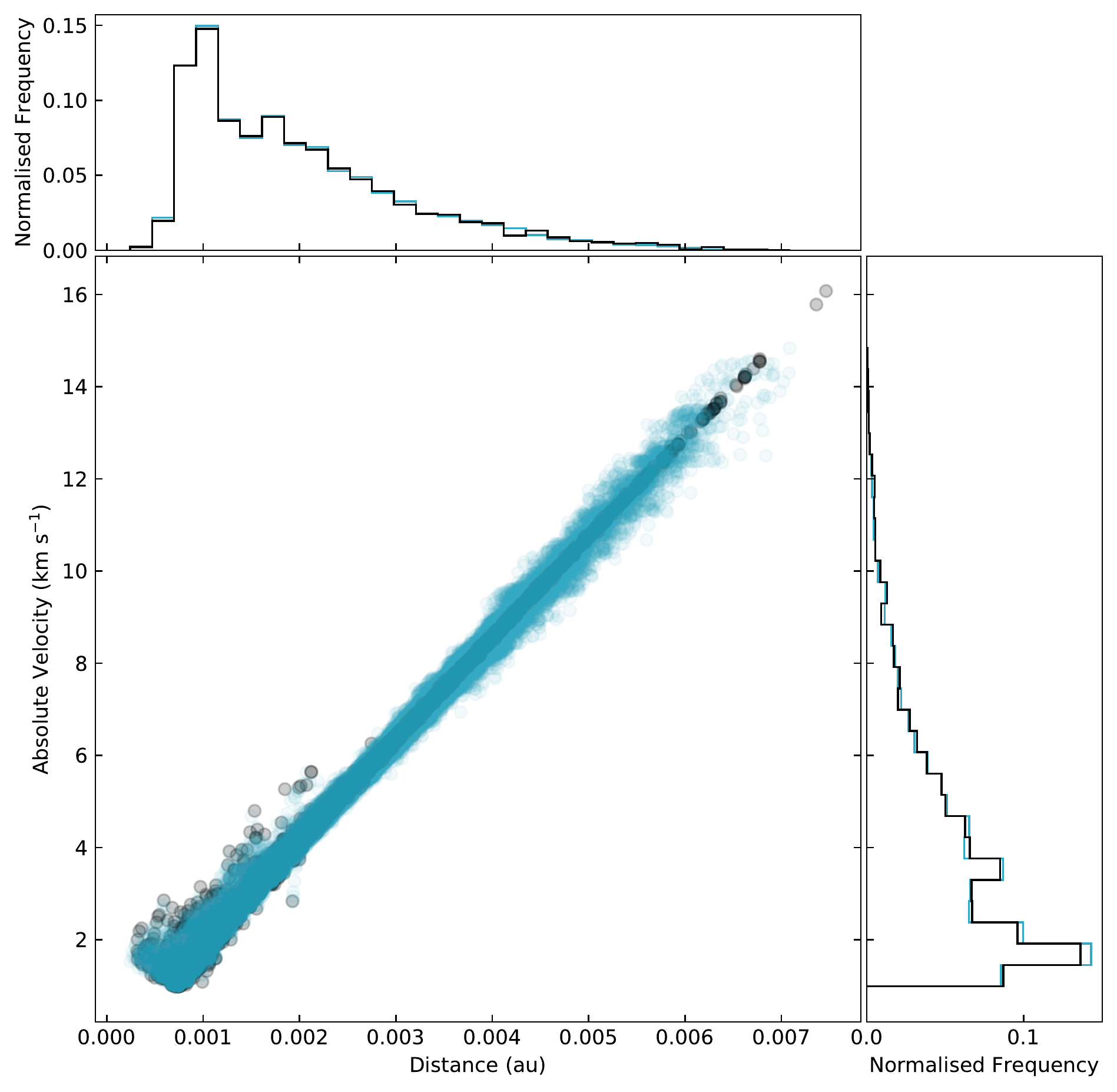}
    \caption{Compares the distribution of absolute velocity versus the distance from the origin for particles escaping the giant impact in Gadget2 simulation (black) and the generated upscaled particles in the $N$-body code (blue). The origin is defined as the centre of mass of the largest and second largest remnants. Data used from sim 8 which is a giant impact between two 0.1 $\rm{M_{\oplus}}$ embryos at an impact velocity of 10 $\rm{km\ s^{-1}}$ and at $b=0$.}
    \label{fig:v_vs_r}
\end{figure}

The use of inverse distance weighting is to interpolate the velocity of generated particles from the initial SPH particles to upscale to the required particle resolution we want for the $N$-body simulations. The velocity of a generated particle, $v(\boldsymbol{x})$, is calculated from,
\begin{equation*}
    v(\mathbf{x}) = \frac{\Sigma_{i=1}^N w_i(\mathbf{x})v_i}{\Sigma_{i=1}^N w_i(\mathbf{x})},
\end{equation*}
where
\begin{equation*}
    w_i(\mathbf{x}) = \frac{1}{d(\mathbf{x},\mathbf{x}_i)^3},
\end{equation*}
here the weight, $w_i(\mathbf{x})$, of the velocity of particle $i$, $v_i$, is calculated from the inverse of the distance between the generated particle and particle $i$ cubed. We choose to cube the distance so the velocity of the generated particle is strongly influenced by the particles close by. $v(\mathbf{x})$ is calculated from the original SPH particle and its 4 closest neighbours. Figure \ref{fig:v_vs_r} shows an example of the generated $N$-body (blue) distribution against the original Gadget 2 (black) distribution of particles for sim 8 in table \ref{tab:collisions_less_1M_e} for the absolute velocity versus the distance from the origin. The origin here is defined as the centre of mass between the largest and second largest remnants post-impact. The figure shows that our generated data matches the original data well.

\subsection{Full SPH Simulation Results}
Below are the full results of the SPH simulations used in this paper. The data has been split into two tables; the table \ref{tab:collisions_less_1M_e} shows results for impacts with a total mass less than $1\ M_{\oplus}$, table \ref{tab:collisions_geq_1M_e} shows impacts with a total mass greater than or equal to $1\ M_{\oplus}$.
\begin{table*}
    \caption{Summary of all parameters and results from SPH simulations for masses below $1\rm{M_\oplus}$. $\rm{M_{tot}}\ -$ total mass in collision in earth masses; $\rm{M_{targ}/M_{proj}}\ -$ mass ratio of target mass to projectile mass; N $-$ number of particles in the simulation; $v_i\ - $ impact velocity in $\rm{km\ s^{-1}}$; $v_i/v_{esc}\ -$  impact velocity normalised by mutual escape velocity; $b -$ impact parameter; $\rm{M_{lr}/M_{tot}}\ -$ ratio of largest remnant mass to total mass; $\rm{M_{unb}/M_{tot}}\ -$ ratio of escaping mass to total mass; $\rm{M_{vap}/M_{tot}\ -}$ ratio of escaping vapour mass to total mass; $Q_R/Q_{RD}^*\ -$ specific impact energy normalised by the catastrophic disruption threshold. \newline \newline
    * $Q_{RD}^*$ is calculated from interpolating $Q_R$ values either side of a giant impact having $M_{lr}$ larger ans smaller than half the total mass. \newline
    $\dagger$ $Q_{RD}^*$ is calculated using methods outlined in  \citealt{Leinhardt-2012-collisions-between-gravity-dom-bodies} using the interpolated $Q_{RD}^*$ value as a base. \newline
    The rest were calculated using methods outlined in \citealt{Leinhardt-2012-collisions-between-gravity-dom-bodies}.}
    \label{tab:collisions_less_1M_e}
    \begin{tabular}{ c c c c c c c c c c c }
    Index & $\rm{M_{tot}}$ & $\rm{M_{proj}/M_{targ}}$ & N & $v_i$ & $v_i/v_{esc}$ & $b$ & $\rm{M_{lr}/M_{tot}}$ & $\rm{M_{unb}/M_{tot}}$ & $\rm{M_{vap}/M_{tot}}$ & $Q_R/Q_{RD}^*$ \\
    $-$ & $\rm{M_{\oplus}}$ & $-$ & $10^4$ &$\rm{km\ s^{-1}}$ & $-$ & $-$ & $-$ & $-$ & $10^{-2}$ & $-$ \\
    \hline\hline
    1 & 0.19 & 0.561 & 6.0 & 5.6 & 1.11 & 0.0 & 0.95 & 0.05 & 0.80 & 0.196 \\
    2 & 0.19 & 0.561 & 6.0 & 6.7 & 1.33 & 0.0 & 0.91 & 0.09 & 1.28 & 0.280 \\
    3 & 0.19 & 0.561 & 6.0 & 7.1 & 1.41 & 0.0 & 0.89 & 0.11 & 1.55 & 0.315 \\
    4 & 0.19 & 0.561 & 6.0 & 8.0 & 1.59 & 0.0 & 0.84 & 0.16 & 2.33 & 0.400 \\
    5 & 0.19 & 0.561 & 6.0 & 8.8 & 1.75 & 0.0 & 0.77 & 0.23 & 2.72 & 0.484 \\
    6 & 0.19 & 0.561 & 6.0 & 9.8 & 1.95 & 0.0 & 0.70 & 0.29 & 3.65 & 0.600 \\
    \hline
    7 & 0.20 & 1.000 & 4.0 & 9.7 & 1.93 & 0.0 & 0.67 & 0.33 & 4.20 & 0.560$^*$ \\
    8 & 0.20 & 1.000 & 4.0 & 10.0 & 1.99 & 0.0 & 0.73 & 0.26 & 3.34 & 0.596$^*$ \\
    9 & 0.20 & 1.000 & 4.0 & 12.0 & 2.38 & 0.0 & 0.55 & 0.44 & 6.45 & 0.858$^*$ \\
    10 & 0.20 & 1.000 & 4.0 & 14.0 & 2.78 & 0.0 & 0.44 & 0.56 & 9.94 & 1.167$^*$ \\
    11 & 0.20 & 1.000 & 4.0 & 15.0 & 2.98 & 0.0 & 0.38 & 0.62 & 12.11 & 1.340$^*$ \\
    \hline
    12 & 0.24 & 1.000 & 20.0 & 5.6 & 1.03 & 0.0 & 0.95 & 0.05 & 0.71 & 0.193 \\
    13 & 0.24 & 1.000 & 20.0 & 6.6 & 1.21 & 0.0 & 0.91 & 0.09 & 1.40 & 0.269 \\
    14 & 0.24 & 1.000 & 20.0 & 7.1 & 1.30 & 0.0 & 0.89 & 0.11 & 1.92 & 0.311 \\
    15 & 0.24 & 1.000 & 20.0 & 7.9 & 1.45 & 0.0 & 0.85 & 0.15 & 2.90 & 0.385 \\
    16 & 0.24 & 1.000 & 20.0 & 8.6 & 1.58 & 0.0 & 0.82 & 0.18 & 2.77 & 0.456 \\
    17 & 0.24 & 1.000 & 20.0 & 9.7 & 1.78 & 0.0 & 0.72 & 0.27 & 3.44 & 0.580 \\
    \hline
    18 & 0.32 & 0.262 & 12.0 & 7.9 & 1.27 & 0.0 & 0.97 & 0.03 & 0.80 & 0.145 \\
    19 & 0.32 & 0.262 & 12.0 & 9.4 & 1.52 & 0.0 & 0.93 & 0.07 & 1.48 & 0.205 \\
    20 & 0.32 & 0.262 & 12.0 & 10.0 & 1.61 & 0.0 & 0.92 & 0.08 & 1.85 & 0.232 \\
    21 & 0.32 & 0.262 & 12.0 & 11.2 & 1.81 & 0.0 & 0.87 & 0.12 & 2.60 & 0.291 \\
    22 & 0.32 & 0.262 & 12.0 & 12.3 & 1.98 & 0.0 & 0.81 & 0.18 & 3.21 & 0.351 \\
    23 & 0.32 & 0.262 & 12.0 & 13.8 & 2.23 & 0.0 & 0.74 & 0.26 & 4.66 & 0.442 \\
    \hline
    24 & 0.38 & 0.466 & 14.0 & 7.3 & 1.13 & 0.0 & 0.95 & 0.05 & 1.30 & 0.187 \\
    25 & 0.38 & 0.466 & 14.0 & 8.7 & 1.35 & 0.0 & 0.92 & 0.08 & 2.08 & 0.265 \\
    26 & 0.38 & 0.466 & 14.0 & 9.3 & 1.45 & 0.0 & 0.89 & 0.11 & 2.99 & 0.303 \\
    27 & 0.38 & 0.358 & 7.0 & 10.0 & 1.54 & 0.0 & 0.93 & 0.07 & 1.41 & 0.255$^*$ \\
    28 & 0.38 & 0.466 & 14.0 & 10.4 & 1.62 & 0.0 & 0.85 & 0.15 & 3.05 & 0.379 \\
    29 & 0.38 & 0.466 & 14.0 & 11.4 & 1.77 & 0.0 & 0.77 & 0.23 & 4.25 & 0.456 \\
    30 & 0.38 & 0.466 & 14.0 & 12.7 & 1.97 & 0.0 & 0.68 & 0.32 & 5.88 & 0.566 \\
    31 & 0.38 & 0.358 & 7.0 & 15.0 & 2.31 & 0.0 & 0.69 & 0.31 & 6.59 & 0.574$^*$ \\
    32 & 0.38 & 0.358 & 7.0 & 17.0 & 2.62 & 0.0 & 0.62 & 0.38 & 8.81 & 0.737$^*$ \\
    33 & 0.38 & 0.358 & 7.0 & 20.0 & 3.08 & 0.0 & 0.49 & 0.51 & 13.57 & 1.020$^*$ \\
    \hline
    34 & 0.56 & 1.000 & 10.0 & 5.2 & 0.70 & 0.0 & 0.96 & 0.04 & 2.37 & 0.095 \\
    35 & 0.56 & 1.000 & 10.0 & 7.4 & 1.00 & 0.0 & 0.94 & 0.06 & 2.41 & 0.193 \\
    36 & 0.56 & 1.000 & 10.0 & 9.0 & 1.21 & 0.0 & 0.89 & 0.10 & 3.38 & 0.286 \\
    37 & 0.56 & 1.000 & 10.0 & 10.0 & 1.35 & 0.0 & 0.93 & 0.07 & 2.10 & 0.353 \\
    38 & 0.56 & 1.000 & 10.0 & 10.4 & 1.40 & 0.0 & 0.85 & 0.15 & 3.66 & 0.381 \\
    39 & 0.56 & 1.000 & 10.0 & 15.0 & 2.02 & 0.0 & 0.69 & 0.31 & 7.49 & 0.793 \\
    \hline
    40 & 0.62 & 0.190 & 12.0 & 10.0 & 1.26 & 0.0 & 0.99 & 0.01 & 0.43 & 0.103 \\
    41 & 0.62 & 0.190 & 12.0 & 15.0 & 1.90 & 0.0 & 0.91 & 0.09 & 2.56 & 0.232 \\
    \hline
    42 & 0.80 & 0.531 & 15.0 & 10.0 & 1.17 & 0.0 & 0.97 & 0.03 & 1.21 & 0.207$^*$ \\
    43 & 0.80 & 0.531 & 15.0 & 15.0 & 1.76 & 0.0 & 0.81 & 0.19 & 5.10 & 0.465$^*$ \\
    44 & 0.80 & 0.531 & 15.0 & 18.9 & 2.22 & 0.0 & 0.64 & 0.36 & 10.85 & 0.738$^*$ \\
    45 & 0.80 & 0.531 & 15.0 & 23.2 & 2.73 & 0.0 & 0.44 & 0.56 & 18.17 & 1.112$^*$ \\
    \hline
    46 & 0.87 & 0.129 & 17.0 & 10.0 & 1.10 & 0.0 & 0.99 & 0.01 & 0.35 & 0.049 \\
    47 & 0.87 & 0.129 & 17.0 & 15.0 & 1.66 & 0.0 & 0.97 & 0.03 & 0.83 & 0.110 \\
    \hline
    \end{tabular}
\end{table*}
\begin{table*}
    \contcaption{}
    \label{tab:continued}
    \begin{tabular}{ c c c c c c c c c c c }
    Index & $\rm{M_{tot}}$ & $\rm{M_{proj}/M_{targ}}$ & N & $v_i$ & $v_i/v_{esc}$ & $b$ & $\rm{M_{lr}/M_{tot}}$ & $\rm{M_{unb}/M_{tot}}$ & $\rm{M_{vap}/M_{tot}}$ & $Q_R/Q_{RD}^*$ \\
    $-$ & $\rm{M_{\oplus}}$ & $-$ & $10^4$ &$\rm{km\ s^{-1}}$ & $-$ & $-$ & $-$ & $-$ & $10^{-2}$ & $-$ \\
    \hline\hline
    48 & 0.19 & 0.561 & 6.0 & 6.3 & 1.25 & 0.4 & 0.93 & 0.07 & 1.01 & 0.182 \\
    49 & 0.19 & 0.561 & 6.0 & 7.5 & 1.49 & 0.4 & 0.88 & 0.12 & 0.93 & 0.258 \\
    50 & 0.19 & 0.561 & 6.0 & 8.0 & 1.59 & 0.4 & 0.62 & 0.13 & 1.08 & 0.293 \\
    51 & 0.19 & 0.561 & 6.0 & 8.9 & 1.77 & 0.4 & 0.59 & 0.17 & 1.45 & 0.363 \\
    52 & 0.19 & 0.561 & 6.0 & 9.8 & 1.95 & 0.4 & 0.57 & 0.21 & 2.00 & 0.440 \\
    53 & 0.19 & 0.561 & 6.0 & 10.9 & 2.17 & 0.4 & 0.53 & 0.26 & 2.70 & 0.545 \\
    \hline
    54 & 0.20 & 1.000 & 4.0 & 10.0 & 1.99 & 0.4 & 0.40 & 0.19 & 1.86 & 0.418$^{\dagger}$ \\
    55 & 0.20 & 1.000 & 4.0 & 15.0 & 2.98 & 0.4 & 0.26 & 0.48 & 7.23 & 0.940$^{\dagger}$ \\
    \hline
    56 & 0.24 & 1.000 & 20.0 & 6.7 & 1.23 & 0.4 & 0.93 & 0.07 & 1.62 & 0.195 \\
    57 & 0.24 & 1.000 & 20.0 & 7.9 & 1.45 & 0.4 & 0.89 & 0.11 & 1.32 & 0.271 \\
    58 & 0.24 & 1.000 & 20.0 & 8.4 & 1.54 & 0.4 & 0.44 & 0.12 & 1.52 & 0.306 \\
    59 & 0.24 & 1.000 & 20.0 & 9.4 & 1.72 & 0.4 & 0.40 & 0.18 & 1.73 & 0.384 \\
    60 & 0.24 & 1.000 & 20.0 & 10.3 & 1.89 & 0.4 & 0.36 & 0.27 & 2.46 & 0.461 \\
    61 & 0.24 & 1.000 & 20.0 & 11.5 & 2.11 & 0.4 & 0.32 & 0.33 & 3.41 & 0.574 \\
    \hline
    62 & 0.32 & 0.262 & 12.0 & 7.7 & 1.24 & 0.4 & 0.95 & 0.05 & 0.68 & 0.113 \\
    63 & 0.32 & 0.262 & 12.0 & 9.1 & 1.47 & 0.4 & 0.92 & 0.08 & 1.25 & 0.158 \\
    64 & 0.32 & 0.262 & 12.0 & 9.7 & 1.56 & 0.4 & 0.90 & 0.10 & 1.71 & 0.179 \\
    65 & 0.32 & 0.262 & 12.0 & 10.9 & 1.76 & 0.4 & 0.80 & 0.11 & 1.55 & 0.226 \\
    66 & 0.32 & 0.262 & 12.0 & 11.9 & 1.92 & 0.4 & 0.76 & 0.14 & 1.99 & 0.270 \\
    67 & 0.32 & 0.262 & 12.0 & 13.3 & 2.15 & 0.4 & 0.73 & 0.17 & 2.71 & 0.337 \\
    \hline
    68 & 0.38 & 0.466 & 14.0 & 8.0 & 1.24 & 0.4 & 0.92 & 0.07 & 1.69 & 0.168 \\
    69 & 0.38 & 0.466 & 14.0 & 9.5 & 1.48 & 0.4 & 0.90 & 0.10 & 1.29 & 0.236 \\
    70 & 0.38 & 0.358 & 7.0 & 10.0 & 1.54 & 0.4 & 0.91 & 0.09 & 1.33 & 0.189$^{\dagger}$ \\
    71 & 0.38 & 0.466 & 14.0 & 10.2 & 1.58 & 0.4 & 0.68 & 0.12 & 1.71 & 0.272 \\
    72 & 0.38 & 0.466 & 14.0 & 11.4 & 1.77 & 0.4 & 0.65 & 0.17 & 2.40 & 0.340 \\
    73 & 0.38 & 0.466 & 14.0 & 12.5 & 1.94 & 0.4 & 0.62 & 0.20 & 3.08 & 0.409 \\
    74 & 0.38 & 0.466 & 14.0 & 13.9 & 2.16 & 0.4 & 0.58 & 0.25 & 4.22 & 0.506 \\
    75 & 0.38 & 0.358 & 7.0 & 15.0 & 2.31 & 0.4 & 0.67 & 0.23 & 3.98 & 0.424$^{\dagger}$ \\
    \hline
    76 & 0.56 & 1.000 & 10.0 & 10.0 & 1.35 & 0.4 & 0.95 & 0.05 & 1.73 & 0.248 \\
    77 & 0.56 & 1.000 & 10.0 & 15.0 & 2.02 & 0.4 & 0.38 & 0.24 & 4.44 & 0.559 \\
    \hline
    78 & 0.62 & 0.190 & 12.0 & 10.0 & 1.26 & 0.4 & 0.97 & 0.03 & 0.43 & 0.085 \\
    79 & 0.62 & 0.190 & 12.0 & 15.0 & 1.90 & 0.4 & 0.90 & 0.10 & 1.94 & 0.191 \\
    \hline
    80 & 0.80 & 0.531 & 15.0 & 10.0 & 1.17 & 0.4 & 0.97 & 0.03 & 0.95 & 0.150$^{\dagger}$ \\
    81 & 0.80 & 0.531 & 15.0 & 15.0 & 1.76 & 0.4 & 0.65 & 0.13 & 2.75 & 0.338$^{\dagger}$ \\
    \hline
    82 & 0.87 & 0.129 & 17.0 & 10.0 & 1.10 & 0.4 & 0.99 & 0.01 & 0.24 & 0.043 \\
    83 & 0.87 & 0.129 & 17.0 & 15.0 & 1.66 & 0.4 & 0.95 & 0.05 & 1.07 & 0.097 \\
    \hline
    84 & 0.19 & 0.561 & 6.0 & 16.9 & 3.36 & 0.8 & 0.60 & 0.10 & 1.16 & 0.146 \\
    85 & 0.19 & 0.561 & 6.0 & 19.9 & 3.96 & 0.8 & 0.59 & 0.12 & 1.74 & 0.203 \\
    86 & 0.19 & 0.561 & 6.0 & 21.3 & 4.24 & 0.8 & 0.59 & 0.14 & 2.04 & 0.233 \\
    87 & 0.19 & 0.561 & 6.0 & 23.8 & 4.73 & 0.8 & 0.58 & 0.16 & 2.64 & 0.291 \\
    88 & 0.19 & 0.561 & 6.0 & 26.1 & 5.19 & 0.8 & 0.57 & 0.18 & 3.30 & 0.349 \\
    89 & 0.19 & 0.561 & 6.0 & 29.2 & 5.81 & 0.8 & 0.55 & 0.21 & 4.30 & 0.437 \\
    \hline
    90 & 0.20 & 1.000 & 4.0 & 10.0 & 1.99 & 0.8 & 0.48 & 0.04 & 0.19 & 0.051$^{\dagger}$ \\
    91 & 0.20 & 1.000 & 4.0 & 15.0 & 2.98 & 0.8 & 0.46 & 0.07 & 0.70 & 0.114$^{\dagger}$ \\
    \hline
    92 & 0.24 & 1.000 & 20.0 & 18.9 & 3.46 & 0.8 & 0.45 & 0.10 & 1.42 & 0.192 \\
    93 & 0.24 & 1.000 & 20.0 & 22.3 & 4.09 & 0.8 & 0.44 & 0.13 & 2.22 & 0.268 \\
    94 & 0.24 & 1.000 & 20.0 & 23.9 & 4.38 & 0.8 & 0.43 & 0.14 & 2.62 & 0.308 \\
    95 & 0.24 & 1.000 & 20.0 & 26.7 & 4.89 & 0.8 & 0.42 & 0.16 & 3.42 & 0.384 \\
    96 & 0.24 & 1.000 & 20.0 & 29.3 & 5.37 & 0.8 & 0.41 & 0.19 & 4.28 & 0.462 \\
    97 & 0.24 & 1.000 & 20.0 & 32.7 & 5.99 & 0.8 & 0.39 & 0.23 & 5.51 & 0.576 \\
    \hline
    98 & 0.32 & 0.262 & 12.0 & 18.7 & 3.02 & 0.8 & 0.77 & 0.09 & 1.23 & 0.072 \\
    99 & 0.32 & 0.262 & 12.0 & 22.1 & 3.56 & 0.8 & 0.77 & 0.10 & 1.76 & 0.100 \\
    100 & 0.32 & 0.262 & 12.0 & 23.6 & 3.81 & 0.8 & 0.76 & 0.11 & 2.03 & 0.114 \\
    101 & 0.32 & 0.262 & 12.0 & 26.4 & 4.26 & 0.8 & 0.76 & 0.13 & 2.62 & 0.143 \\
    102 & 0.32 & 0.262 & 12.0 & 28.9 & 4.66 & 0.8 & 0.75 & 0.15 & 3.22 & 0.172 \\
    103 & 0.32 & 0.262 & 12.0 & 32.3 & 5.21 & 0.8 & 0.74 & 0.17 & 4.15 & 0.214 \\
    \hline
    \end{tabular}
    \end{table*}
    \begin{table*}
    \contcaption{}
    \label{tab:cont2}
    \begin{tabular}{ c c c c c c c c c c c }
    Index & $\rm{M_{tot}}$ & $\rm{M_{proj}/M_{targ}}$ & N & $v_i$ & $v_i/v_{esc}$ & $b$ & $\rm{M_{lr}/M_{tot}}$ & $\rm{M_{unb}/M_{tot}}$ & $\rm{M_{vap}/M_{tot}}$ & $Q_R/Q_{RD}^*$ \\
    $-$ & $\rm{M_{\oplus}}$ & $-$ & $10^4$ &$\rm{km\ s^{-1}}$ & $-$ & $-$ & $-$ & $-$ & $10^{-2}$ & $-$ \\
    \hline\hline
    104 & 0.38 & 0.358 & 7.0 & 10.0 & 1.54 & 0.8 & 0.75 & 0.02 & 0.15 & 0.019$^{\dagger}$ \\
    105 & 0.38 & 0.358 & 7.0 & 15.0 & 2.31 & 0.8 & 0.73 & 0.05 & 0.59 & 0.043$^{\dagger}$ \\
    106 & 0.38 & 0.466 & 14.0 & 21.1 & 3.28 & 0.8 & 0.65 & 0.10 & 1.71 & 0.127 \\
    107 & 0.38 & 0.466 & 14.0 & 24.9 & 3.87 & 0.8 & 0.64 & 0.12 & 2.48 & 0.177 \\
    108 & 0.38 & 0.466 & 14.0 & 26.6 & 4.13 & 0.8 & 0.64 & 0.13 & 2.87 & 0.202 \\
    109 & 0.38 & 0.466 & 14.0 & 29.8 & 4.63 & 0.8 & 0.63 & 0.15 & 3.67 & 0.254 \\
    110 & 0.38 & 0.466 & 14.0 & 32.6 & 5.07 & 0.8 & 0.62 & 0.18 & 4.63 & 0.304 \\
    \hline
    111 & 0.56 & 1.000 & 10.0 & 10.0 & 1.35 & 0.8 & 0.50 & 0.01 & 0.10 & 0.031 \\
    112 & 0.56 & 1.000 & 10.0 & 15.0 & 2.02 & 0.8 & 0.48 & 0.04 & 0.51 & 0.069 \\
    \hline
    113 & 0.62 & 0.190 & 12.0 & 10.0 & 1.26 & 0.8 & 0.87 & 0.01 & 0.13 & 0.009 \\
    114 & 0.62 & 0.190 & 12.0 & 15.0 & 1.90 & 0.8 & 0.85 & 0.03 & 0.46 & 0.020 \\
    115 & 0.62 & 0.190 & 12.0 & 16.5 & 2.09 & 0.8 & 0.84 & 0.06 & 0.98 & 0.024 \\
    116 & 0.62 & 0.190 & 12.0 & 23.4 & 2.96 & 0.8 & 0.83 & 0.09 & 1.84 & 0.048 \\
    117 & 0.62 & 0.190 & 12.0 & 28.6 & 3.61 & 0.8 & 0.82 & 0.11 & 2.67 & 0.072 \\
    118 & 0.62 & 0.190 & 12.0 & 33.0 & 4.17 & 0.8 & 0.81 & 0.13 & 3.53 & 0.096 \\
    119 & 0.62 & 0.190 & 12.0 & 36.9 & 4.66 & 0.8 & 0.80 & 0.15 & 4.50 & 0.119 \\
    \hline
    120 & 0.80 & 0.531 & 15.0 & 10.0 & 1.17 & 0.8 & 0.67 & 0.00 & 0.07 & 0.016$^{\dagger}$ \\
    121 & 0.80 & 0.531 & 15.0 & 15.0 & 1.76 & 0.8 & 0.65 & 0.03 & 0.38 & 0.036$^{\dagger}$ \\
    122 & 0.80 & 0.531 & 15.0 & 19.5 & 2.29 & 0.8 & 0.63 & 0.06 & 1.26 & 0.061$^{\dagger}$ \\
    123 & 0.80 & 0.531 & 15.0 & 27.6 & 3.24 & 0.8 & 0.62 & 0.09 & 2.58 & 0.123$^{\dagger}$ \\
    124 & 0.80 & 0.531 & 15.0 & 33.8 & 3.97 & 0.8 & 0.61 & 0.12 & 3.94 & 0.184$^{\dagger}$ \\
    125 & 0.80 & 0.531 & 15.0 & 39.0 & 4.58 & 0.8 & 0.59 & 0.16 & 5.54 & 0.245$^{\dagger}$ \\
    \hline
    126 & 0.87 & 0.129 & 17.0 & 10.0 & 1.10 & 0.8 & 0.93 & 0.01 & 0.19 & 0.005 \\
    127 & 0.87 & 0.129 & 17.0 & 15.0 & 1.66 & 0.8 & 0.90 & 0.04 & 0.49 & 0.010 \\
    \hline
    \hline \\ 
    \end{tabular}
\end{table*}

\begin{table*}
    \centering
    \caption{Summary of all parameters and results from SPH simulations for masses equal to or greater than $1\rm{M_\oplus}$}
    \label{tab:collisions_geq_1M_e}
    \begin{tabular}{ c c c c c c c c c c c }
    Index & $\rm{M_{tot}}$ & $\rm{M_{proj}/M_{targ}}$ & N & $v_i$ & $v_i/v_{esc}$ & $b$ & $\rm{M_{lr}/M_{tot}}$ & $\rm{M_{unb}/M_{tot}}$ & $\rm{M_{vap}/M_{tot}}$ & $Q_R/Q_{RD}^*$ \\
    $-$ & $\rm{M_{\oplus}}$ & $-$ & $10^4$ &$\rm{km\ s^{-1}}$ & $-$ & $-$ & $-$ & $-$ & $10^{-2}$ & $-$ \\
    \hline\hline
    128 & 1.05 & 1.000 & 20.0 & 6.4 & 0.69 & 0.0 & 0.96 & 0.04 & 2.58 & 0.082$^*$ \\
    129 & 1.05 & 1.000 & 20.0 & 9.1 & 0.98 & 0.0 & 0.94 & 0.06 & 2.87 & 0.165$^*$ \\
    130 & 1.05 & 1.000 & 20.0 & 10.0 & 1.07 & 0.0 & 0.97 & 0.03 & 1.40 & 0.200$^*$ \\
    131 & 1.05 & 0.360 & 20.0 & 10.0 & 1.06 & 0.0 & 0.98 & 0.02 & 1.28 & 0.145 \\
    132 & 1.05 & 1.000 & 20.0 & 11.2 & 1.20 & 0.0 & 0.89 & 0.11 & 4.16 & 0.251$^*$ \\
    133 & 1.05 & 1.000 & 20.0 & 12.9 & 1.38 & 0.0 & 0.84 & 0.16 & 5.38 & 0.333$^*$ \\
    134 & 1.05 & 0.360 & 20.0 & 15.0 & 1.59 & 0.0 & 0.91 & 0.09 & 3.14 & 0.327 \\
    135 & 1.05 & 1.000 & 20.0 & 15.0 & 1.61 & 0.0 & 0.83 & 0.17 & 5.41 & 0.450$^*$ \\
    136 & 1.05 & 1.000 & 20.0 & 18.6 & 1.99 & 0.0 & 0.67 & 0.33 & 10.12 & 0.691$^*$ \\
    137 & 1.05 & 1.000 & 20.0 & 22.8 & 2.45 & 0.0 & 0.48 & 0.52 & 18.32 & 1.039$^*$ \\
    138 & 1.05 & 0.360 & 20.0 & 24.7 & 2.61 & 0.0 & 0.58 & 0.42 & 14.41 & 0.887 \\
    \hline
    139 & 1.21 & 0.090 & 22.0 & 10.0 & 0.97 & 0.0 & 1.00 & 0.00 & 0.21 & 0.023 \\
    140 & 1.21 & 0.090 & 22.0 & 15.0 & 1.46 & 0.0 & 0.99 & 0.01 & 0.37 & 0.051 \\
    \hline
    141 & 1.29 & 0.679 & 25.0 & 10.0 & 0.99 & 0.0 & 0.97 & 0.03 & 1.79 & 0.159$^*$ \\
    142 & 1.29 & 0.679 & 25.0 & 15.0 & 1.49 & 0.0 & 0.89 & 0.11 & 4.31 & 0.358$^*$ \\
    143 & 1.29 & 0.679 & 25.0 & 25.4 & 2.52 & 0.0 & 0.48 & 0.51 & 18.47 & 1.026$^*$ \\
    \hline
    144 & 1.39 & 0.250 & 25.0 & 10.0 & 0.95 & 0.0 & 0.99 & 0.01 & 0.58 & 0.084 \\
    145 & 1.39 & 0.250 & 25.0 & 15.0 & 1.42 & 0.0 & 0.96 & 0.04 & 1.64 & 0.188 \\
    \hline
    146 & 1.54 & 1.000 & 30.0 & 10.0 & 0.93 & 0.0 & 0.97 & 0.03 & 1.96 & 0.152$^*$ \\
    147 & 1.54 & 1.000 & 30.0 & 15.0 & 1.40 & 0.0 & 0.90 & 0.10 & 4.34 & 0.341$^*$ \\
    148 & 1.54 & 1.000 & 30.0 & 21.2 & 1.97 & 0.0 & 0.67 & 0.33 & 11.35 & 0.681$^*$ \\
    149 & 1.54 & 1.000 & 30.0 & 25.9 & 2.41 & 0.0 & 0.49 & 0.51 & 19.20 & 1.017$^*$ \\
    \hline
    150 & 1.63 & 0.471 & 30.0 & 10.0 & 0.91 & 0.0 & 0.98 & 0.02 & 1.43 & 0.133 \\
    151 & 1.63 & 0.471 & 30.0 & 15.0 & 1.36 & 0.0 & 0.93 & 0.07 & 2.92 & 0.299 \\
    152 & 1.63 & 0.471 & 30.0 & 25.1 & 2.28 & 0.0 & 0.62 & 0.38 & 13.44 & 0.837 \\
    \hline
    153 & 1.88 & 0.694 & 35.0 & 10.0 & 0.87 & 0.0 & 0.97 & 0.03 & 2.23 & 0.136$^*$ \\
    154 & 1.88 & 0.694 & 35.0 & 23.4 & 2.03 & 0.0 & 0.65 & 0.35 & 12.77 & 0.744$^*$ \\
    155 & 1.88 & 0.694 & 35.0 & 28.7 & 2.49 & 0.0 & 0.43 & 0.57 & 21.83 & 1.119$^*$ \\
    \hline
    156 & 2.22 & 1.000 & 40.0 & 23.9 & 1.96 & 0.0 & 0.66 & 0.34 & 12.70 & 0.799 \\
    \hline
    157 & 1.05 & 1.000 & 20.0 & 10.0 & 1.07 & 0.4 & 0.98 & 0.02 & 1.09 & 0.140$^{\dagger}$ \\
    158 & 1.05 & 0.360 & 20.0 & 10.0 & 1.06 & 0.4 & 0.98 & 0.02 & 0.60 & 0.111 \\
    159 & 1.05 & 0.360 & 20.0 & 15.0 & 1.59 & 0.4 & 0.91 & 0.09 & 2.36 & 0.249 \\
    160 & 1.05 & 1.000 & 20.0 & 15.0 & 1.61 & 0.4 & 0.45 & 0.10 & 2.36 & 0.315$^{\dagger}$ \\
    \hline
    161 & 1.21 & 0.090 & 22.0 & 10.0 & 0.97 & 0.4 & 1.00 & 0.00 & 0.12 & 0.021 \\
    162 & 1.21 & 0.090 & 22.0 & 15.0 & 1.46 & 0.4 & 0.98 & 0.02 & 0.63 & 0.048 \\
    \hline
    163 & 1.29 & 0.679 & 25.0 & 10.0 & 0.99 & 0.4 & 0.98 & 0.02 & 0.99 & 0.113$^{\dagger}$ \\
    164 & 1.29 & 0.679 & 25.0 & 15.0 & 1.49 & 0.4 & 0.90 & 0.10 & 3.59 & 0.254$^{\dagger}$ \\
    \hline
    165 & 1.39 & 0.250 & 25.0 & 10.0 & 0.95 & 0.4 & 0.99 & 0.01 & 0.40 & 0.068 \\
    166 & 1.39 & 0.250 & 25.0 & 15.0 & 1.42 & 0.4 & 0.94 & 0.06 & 1.64 & 0.153 \\
    \hline
    167 & 1.54 & 1.000 & 30.0 & 10.0 & 0.93 & 0.4 & 0.99 & 0.01 & 1.05 & 0.106$^{\dagger}$ \\
    168 & 1.54 & 1.000 & 30.0 & 15.0 & 1.40 & 0.4 & 0.94 & 0.06 & 2.88 & 0.239$^{\dagger}$ \\
    \hline
    169 & 1.63 & 0.471 & 30.0 & 10.0 & 0.91 & 0.4 & 0.99 & 0.01 & 0.67 & 0.098 \\
    170 & 1.63 & 0.471 & 30.0 & 15.0 & 1.36 & 0.4 & 0.93 & 0.07 & 2.47 & 0.221 \\
    \hline
    171 & 1.88 & 0.694 & 35.0 & 10.0 & 0.87 & 0.4 & 0.99 & 0.01 & 0.89 & 0.097$^{\dagger}$ \\
    172 & 1.88 & 0.694 & 35.0 & 15.0 & 1.30 & 0.4 & 0.95 & 0.05 & 2.21 & 0.218$^{\dagger}$ \\
    \hline
    173 & 2.22 & 1.000 & 40.0 & 10.0 & 0.82 & 0.4 & 0.99 & 0.01 & 1.07 & 0.099 \\
    174 & 2.22 & 1.000 & 40.0 & 15.0 & 1.23 & 0.4 & 0.97 & 0.03 & 1.82 & 0.222 \\
    \hline
    175 & 1.05 & 1.000 & 20.0 & 10.0 & 1.07 & 0.8 & 0.50 & 0.00 & 0.04 & 0.017$^{\dagger}$ \\
    176 & 1.05 & 0.360 & 20.0 & 10.0 & 1.06 & 0.8 & 1.00 & 0.00 & 0.09 & 0.012 \\
    177 & 1.05 & 0.360 & 20.0 & 15.0 & 1.59 & 0.8 & 0.74 & 0.02 & 0.35 & 0.026 \\
    178 & 1.05 & 1.000 & 20.0 & 15.0 & 1.61 & 0.8 & 0.49 & 0.02 & 0.33 & 0.038$^{\dagger}$ \\
    \hline
    \end{tabular}
\end{table*}
\begin{table*}
    \contcaption{}
    \label{tab:table2cont}
    \begin{tabular}{ c c c c c c c c c c c }
    Index & $\rm{M_{tot}}$ & $\rm{M_{proj}/M_{targ}}$ & N & $v_i$ & $v_i/v_{esc}$ & $b$ & $\rm{M_{lr}/M_{tot}}$ & $\rm{M_{unb}/M_{tot}}$ & $\rm{M_{vap}/M_{tot}}$ & $Q_R/Q_{RD}^*$ \\
    $-$ & $\rm{M_{\oplus}}$ & $-$ & $10^4$ &$\rm{km\ s^{-1}}$ & $-$ & $-$ & $-$ & $-$ & $10^{-2}$ & $-$ \\
    \hline\hline
    179 & 1.21 & 0.090 & 22.0 & 10.0 & 0.97 & 0.8 & 0.98 & 0.01 & 0.22 & 0.002 \\
    180 & 1.21 & 0.090 & 22.0 & 15.0 & 1.46 & 0.8 & 0.93 & 0.03 & 0.47 & 0.006 \\
    181 & 1.21 & 0.090 & 22.0 & 18.3 & 1.78 & 0.8 & 0.92 & 0.06 & 1.20 & 0.008 \\
    182 & 1.21 & 0.090 & 22.0 & 25.8 & 2.50 & 0.8 & 0.91 & 0.08 & 1.87 & 0.016 \\
    183 & 1.21 & 0.090 & 22.0 & 31.7 & 3.08 & 0.8 & 0.91 & 0.09 & 2.43 & 0.025 \\
    \hline
    184 & 1.29 & 0.679 & 25.0 & 10.0 & 0.99 & 0.8 & 0.99 & 0.01 & 0.64 & 0.013$^{\dagger}$ \\
    185 & 1.29 & 0.679 & 25.0 & 15.0 & 1.49 & 0.8 & 0.60 & 0.02 & 0.30 & 0.028$^{\dagger}$ \\
    \hline
    186 & 1.39 & 0.250 & 25.0 & 10.0 & 0.95 & 0.8 & 0.99 & 0.01 & 0.35 & 0.007 \\
    187 & 1.39 & 0.250 & 25.0 & 15.0 & 1.42 & 0.8 & 0.81 & 0.02 & 0.36 & 0.016 \\
    \hline
    188 & 1.54 & 1.000 & 30.0 & 10.0 & 0.93 & 0.8 & 0.99 & 0.01 & 0.97 & 0.013$^{\dagger}$ \\
    189 & 1.54 & 1.000 & 30.0 & 15.0 & 1.40 & 0.8 & 0.49 & 0.01 & 0.27 & 0.029$^{\dagger}$ \\
    \hline
    190 & 1.63 & 0.471 & 30.0 & 10.0 & 0.91 & 0.8 & 0.99 & 0.01 & 0.44 & 0.011 \\
    191 & 1.63 & 0.471 & 30.0 & 15.0 & 1.36 & 0.8 & 0.69 & 0.01 & 0.29 & 0.024 \\
    \hline
    192 & 1.88 & 0.694 & 35.0 & 10.0 & 0.87 & 0.8 & 0.99 & 0.01 & 0.49 & 0.011$^{\dagger}$ \\
    193 & 1.88 & 0.694 & 35.0 & 15.0 & 1.30 & 0.8 & 0.59 & 0.01 & 0.27 & 0.025$^{\dagger}$ \\
    \hline
    194 & 2.22 & 1.000 & 40.0 & 10.0 & 0.82 & 0.8 & 0.99 & 0.01 & 0.52 & 0.012 \\
    195 & 2.22 & 1.000 & 40.0 & 15.0 & 1.23 & 0.8 & 0.50 & 0.01 & 0.22 & 0.028 \\
    \hline
    \hline \\ 
    \end{tabular}
\end{table*}





\bsp	
\label{lastpage}
\end{document}